\documentclass{cernrep} 
\usepackage{texnames}
\usepackage{epsfig,axodraw}
\usepackage{amssymb}
\usepackage{graphicx}
\usepackage{bm}

\usepackage[T1]{fontenc}
\newcommand{\qs}{Q_{\rm s, p}}
\newcommand{\qsh}{Q_{\rm s, h}}
\newcommand{\qsa}{Q_{\rm s, A}}

\begin{document}
\title{Lectures on high-energy heavy-ion collisions at the LHC\footnote{Lectures given at {\it The 2008 European School of High-Energy Physics}, Herbeumont-sur-Semois, Belgium 8 --- 21 June 2008}}
 
\author{Carlos A. Salgado\footnote{E-mail address: carlos.salgado@usc.es}}

\institute{Departamento de F\'\i sica de Part\'\i culas and IGFAE\\Universidade de Santiago de Compostela\\E-15782 Santiago de Compostela (Galicia-Spain)}

\maketitle 

\begin{abstract}
Some topics on heavy-ion collisions are reviewed with emphasis on those which are expected to be specially relevant at the Large Hadron Collider programme. 
\end{abstract}
 
\section{Introduction}
 
Heavy ions will be collided at the LHC with energies thirty times larger than ever before: each individual nucleon inside the colliding nuclei will be accelerated at three times the top Tevatron energy in the centre-of-mass frame --- which is already ten times more than that in AuAu collisions at the Relativistic Heavy Ion Collider (RHIC) in Brookhaven. The energy frontier reached in heavy-ion  and proton--proton collisions is, for the first time, the same, since the same machine is used to accelerate the two different systems\footnote{There is a small difference in the energy per nucleon in different colliding systems due to the mass-over-charge ratio being larger for larger nuclei. For this reason, the energy for each of the accelerated nuclei at the LHC will be $Z/A\times$7 GeV, with $A$ the atomic number and $Z$ the number of protons.}. Most of the currently targeted physics quests refer, however, to very different energy scales; those in heavy ions correspond to typical QCD scales ($\Lambda_{\rm QCD}<$ 1 GeV) while the scales for the principal LHC searches in the proton--proton programme lie 2--3 orders of magnitude above.

The main goal of heavy-ion collisions is to form and characterize a macroscopic (in QCD scales) state of deconfined quarks and gluons in local thermal equilibrium. The large amount of energy liberated in the collision is distributed over distances of the transverse size of an atomic nuclei ($R_A\sim A^{1/3}$ fm $\sim 6$ fm for a lead nucleus) and expected to give rise to collective effects with distinctive signatures. This distribution of energy in the largest possible length scale is the most probable configuration in any hadronic collision. On the other hand, the configuration in which a large amount of energy is concentrated in a small region of phase space occurs very rarely, but interestingly leads to the largest probability of producing something unknown like a new particle. The second possibility is specially relevant for the searches in the LHC proton--proton programme. With the study of hard processes it is also becoming more and more important for the heavy-ion programme.

In some cases, a nucleus--nucleus collision can be considered as a superposition of independent proton--proton collisions. This happens, for example, for hard enough probes, as photon production, in which the scales probed are much larger than those ruling the collective behaviour (or, in other words, those for which the production time is much shorter than the typical times involved in collective behaviour). These probes are very important as benchmarks. In general, however, the output of a heavy-ion collision is very different from such a simple superposition. The study of processes sensitive to the degree of collectivity of the system is the main goal of heavy-ion collisions. For example, a powerful way of looking for collective behaviour is by studying the azimuthal distribution of particles produced at a given transverse momentum $p_T$: if the collisions were independent, the particles would be uniformly distributed (within statistical fluctuations) while the presence of azimuthal asymmetries would indicate that some non-trivial phenomenon is taking place.  Different other {\it probes} of this collective behaviour have been proposed both in the {\it soft } and the {\it hard} sector of QCD --- ruled by non-perturbative and perturbative dynamics, respectively. In the next sections we shall review some of these probes and the new opportunities opened at the LHC. 

In parallel to the PbPb programme, pPb collisions will also be made at the LHC. One of the main goals will be in this case to provide the essential benchmark for a correct interpretation of the hot matter probes --- the energy densities reached in a pPb collision at the LHC are not expected to form a hot medium in the final state. For example, the knowledge of the parton distribution functions is deficient at small-$x_{\rm Bj}$, those most relevant for the LHC, and especially for the case of gluons. Moreover, this type of collisions present an interesting physics case {\it per se}, where non-linear terms in the evolution equations are expected to become important, marking the beginning of a new regime of QCD where the parton distributions are saturated. Extra physics opportunities are the ultraperipheral collisions and the measurement of cross-sections of cosmic-ray interest.

 \section{The LHC heavy-ion collider and the experiments}
 
 At the LHC, the centre-of-mass energy of a collision of two different systems with charge $Z_1,\,Z_2$ and atomic numbers $A_1,\,A_2$ with $Z=A=1$ for a proton is (see Ref. \cite{Jowett:2008hb} for an updated report on the subject of nuclear collisions at the LHC)
 \begin{equation}
 \sqrt{s_{NN}}\simeq 2\sqrt{s_{pp}}+\sqrt{s_{pp}}\sqrt{\frac{Z_1Z_2}{A_1A_2}},
 \label{eq:sqrtAA}
 \end{equation}
where $NN$ in the subindex refers to the energy per nucleon inside the colliding nucleus and $\sqrt{s_{pp}}$ is the corresponding energy in pp collisions. This gives the maximum energies
\begin{eqnarray}
\sqrt{s_{NN}}&\simeq& 5.5\, \mbox{GeV\, for PbPb collisions},\nonumber \\
\sqrt{s_{NN}}&\simeq& 8.8\, \mbox{GeV\, for pPb collisions}.\nonumber 
 \end{eqnarray}
 Additionally, for non-symmetric systems --- as in the projected pA collisions --- there is a rapidity shift
 \begin{equation}
 \Delta y\simeq\frac{1}{2}\log\frac{Z_1A_2}{Z_2A_1}
 \label{eq:rapshift}
 \end{equation}
 due to the fact that the centre-of-mass frame of the pA collision does not coincide with the laboratory centre-of-mass frame. This rapidity shift is $\Delta y\simeq 1$ for proton--lead collisions and would need to be taken into account for the comparison with PbPb data.
 %
\begin{figure}[htbp]
\begin{minipage}{0.5\textwidth}
\center
\includegraphics[scale=0.4]{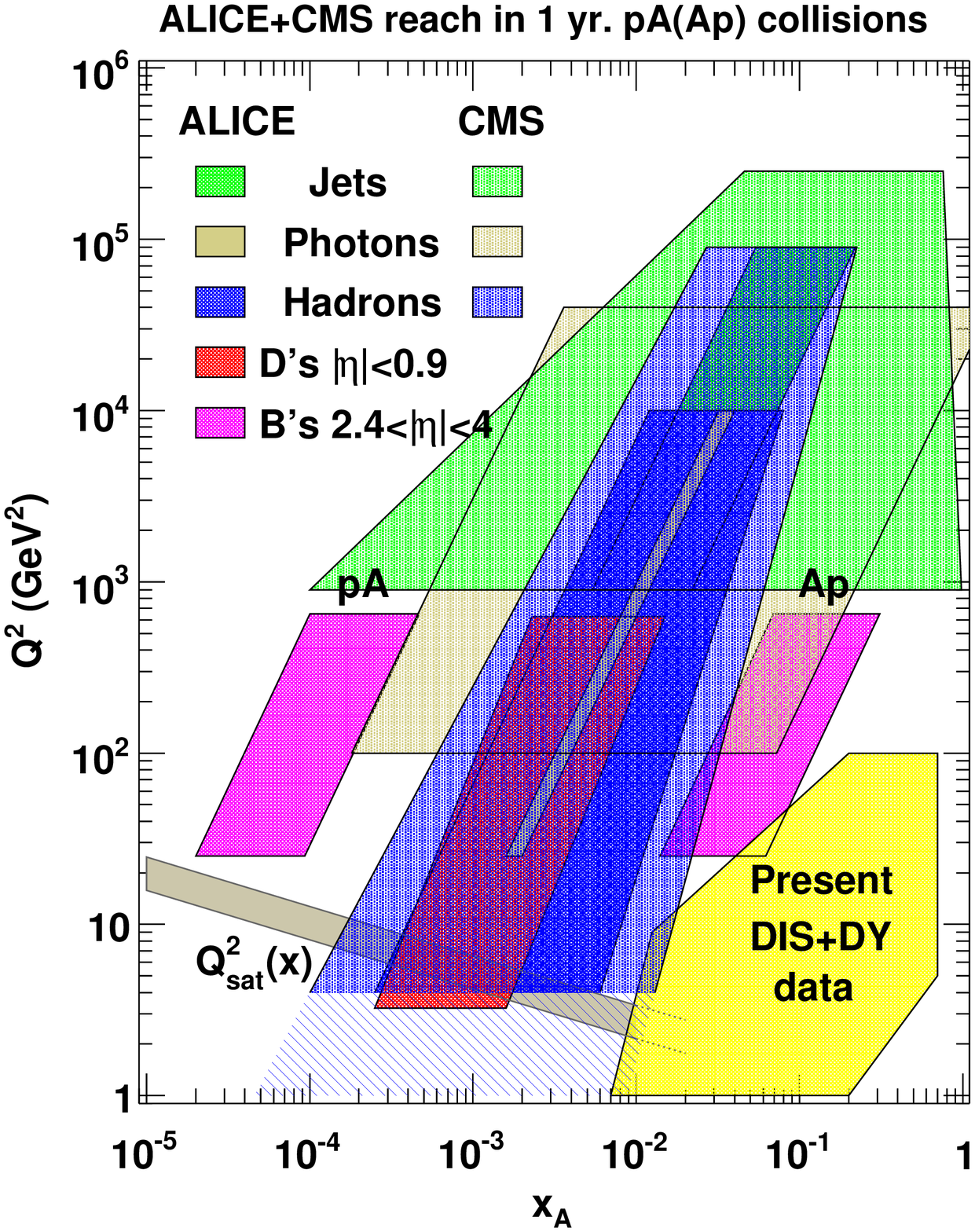}
\end{minipage}
\begin{minipage}{0.5\textwidth}
\center
\includegraphics[scale=0.4]{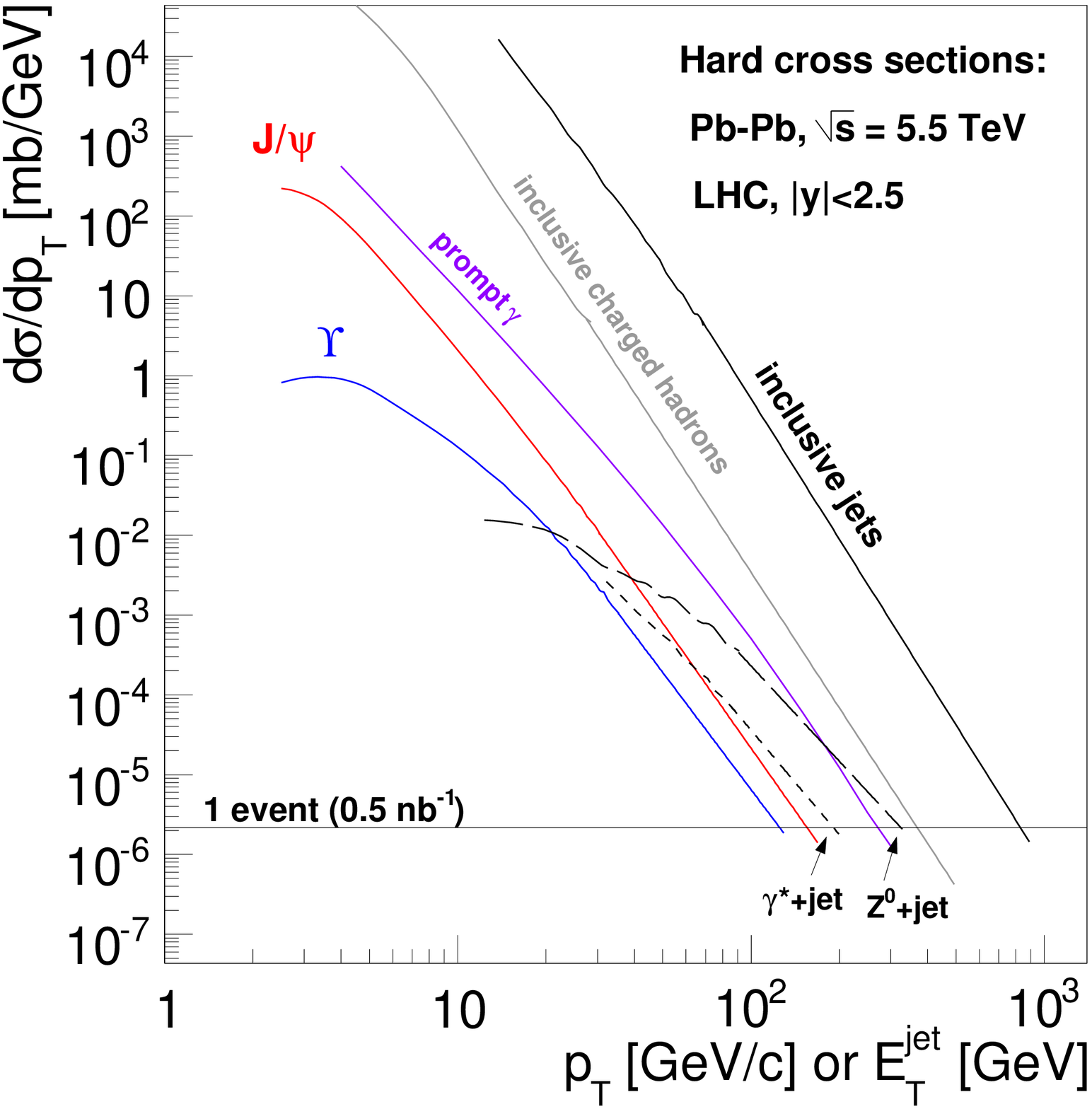}
\end{minipage}
\caption[]{\small (Left) Kinematical reach in one year of pPb collisions at the LHC. PbPb collisions will have a similar reach. (Right) Reach in transverse momentum (energy) for different processes in PbPb collisions at the LHC, figure from Ref. \protect\cite{d'Enterria:2008ge}. }
\label{fig:lhcreach}
\end{figure}

Another relevant quantity is the maximum luminosity for the heavy-ion programme which is estimated to be 
\begin{equation}
L_{NN}\simeq 10^{27}\,  {\rm cm}^{-2}{\rm s}^{-1}
\label{eq:luminosity}
\end{equation}
where $NN$ refers again to luminosity {\it per nucleon}. This luminosity per nucleon is $\sim$ 7 orders of magnitude smaller than the corresponding nominal luminosity for a typical pp run --- $L_{pp}\simeq 10^{34}\,{\rm cm}^{-2}{\rm s}^{-1}$. A typical year when the LHC reaches its design features will be eight months of pp run and one month dedicated to the heavy-ion programme (including nucleus--nucleus or proton--nucleus collisions at different energies or pp at smaller energies for benchmarking in the PbPb runs). This reduces the integrated luminosity by another factor of ten with respect to the proton--proton programme. 

Heavy-ion collisions will be detected by ALICE \cite{Carminati:2004fp,Alessandro:2006yt} --- the LHC experiment dedicate to that purpose --- and both CMS \cite{D'Enterria:2007xr} and ATLAS \cite{atlas}. The three experiments have complementary capabilities for the study of the different probes proposed. They are, for the same reason, flexible enough to look for unknown or not yet considered processes. 

Kinematical reaches are, at the LHC, orders of magnitude different than the present ones both both for large virtualities and small-$x$ --- see Fig. \ref{fig:lhcreach} for the kinematical reach explored in a typical year of pPb collisions. With these numbers it becomes clear that the LHC will provide exploration of completely new regimes of QCD under extreme conditions.

\section{QCD at high temperatures}

QCD is the theory of strong interactions: it describes the interactions between hadrons (protons, $\pi$'s, etc.) which are the asymptotic states of the theory at small temperatures and densities. They form nuclear matter, which constitutes the majority of the observable mass of the Universe. Hadrons are colourless objects. The Lagrangian of QCD is, however, written in terms of quarks and gluons which are the building blocks of the hadrons --- hadrons are composite particles. Quarks and gluons are colourful objects. Colour is the {\it charge} of QCD and the fact that the bosons carrying the interaction, the gluons, are also coloured allows them to interact. This makes the theory very different from its abelian counterpart, QED. Quarks are colour triplets corresponding to the fundamental representation of the QCD gauge theory group $SU(3)$. There are six flavours of quarks with fractional electric charges and very different masses:

\begin{center}
\begin{tabular}{|c|c|c|c|}\hline
charge\,=\,2/3 & $u$ ($\sim$ 5 MeV)& $c$ ($\sim$ 1.5 GeV)& $t$ ($\sim$ 175 GeV)\\\hline
charge\,=\, -1/3 & $d$ ($\sim$ 10 MeV)& $s$ ($\sim$ 100 MeV)& $b$ ($\sim$ 5 GeV)\\
\hline
\end{tabular}
\end{center}

One normally distinguishes between light ($u,\,d,\,s$) and heavy ($c,\,b,\,t$) quarks. Two extremely important properties of QCD are confinement and asymptotic freedom. The first corresponds to the non-existence of colourful asymptotic states --- i.e., colour is confined in regions smaller than a typical QCD scale, $\Lambda_{\rm QCD}^{-1}\sim 1$ fm. In particular, no free quarks or gluons can exist as asymptotic states. Asymptotic freedom, on the other hand, corresponds to the behaviour of the theory at small distances or large scales. The QCD coupling constant becomes smaller and smaller --- the interaction disappears --- at these large scales or small distances \cite{Gross:1973id,Politzer:1973fx}.

A simple picture for the confinement is that of a string with a quark and an antiquark at each end. The potential between the $q\bar q$ pair can be written as \cite{Eichten:1979ms} (this is known as the Cornell potential):
\begin{equation}
V(r)\simeq -\frac{\alpha_{\rm eff}}{r}+\sigma r\, .
\label{eq:pot}
\end{equation}
The linear part of the potential becomes relevant at large distances; $\sigma$ is called the string tension. So, if we try to pull apart the quark from the antiquark, the linear potential will make the amount of energy grow indefinitely; an isolated quark in the vacuum has infinite energy. This, however, does not actually happen since when the energy in the string is larger than $m_q+m_{\bar q}$, the string can break by pair creation and form two different strings --- two different hadrons. In the limit $m_q\to\infty$ the string cannot break.

Another important property of QCD is {\it chiral symmetry}. In the absence of quark masses, the QCD Lagrangian splits into two independent quark sectors,
\begin{equation}
{\cal L}_{QCD}={\cal L}_{\rm gluons}+i\bar q_L\gamma^\mu D_\mu q_L+i\bar q_R\gamma^\mu D_\mu q_R\, .
\label{eq:qcdlagrangian}
\end{equation}
This Lagrangian is symmetric under $SU(N)_L\times SU(N)_R$ with $N$ the number of massless quarks. However, this symmetry is not observed experimentally. The mechanism of spontaneous symmetry breaking does not need in QCD the introduction of a new field, at variance with the electro-weak sector, whose symmetry is broken by the presence of a Higgs field. In  QCD, the vacuum $|0\rangle$ is not invariant under the transformation, the {\it chiral condensate} is non-zero
\begin{equation}
\langle0|\bar q_L q_R+\bar q_R q_L|0\rangle\neq 0.
\end{equation}
The Goldstone theorem states that, when continuous symmetries are spontaneously broken, associated massless bosons should appear. In the case of $N=2$ this corresponds in QCD to the three pions $\pi^\pm$, $\pi^0$ whose masses are much smaller than other typical hadronic masses. The case for $N=3$ includes the lowest-mass strange mesons.

So, two main properties of the QCD vacuum are confinement and chiral symmetry breaking. A relevant question is then: {\it Is there a regime where these symmetries are restored?}. Intuitively, asymptotic freedom should lead to a deconfined free gas of quarks and gluons at very large temperatures and, indeed, this was proposed soon after asymptotic freedom was discovered \cite{qgp} \footnote{The original proposals of a different, deconfined, phase of hadronic matter at high temperatures is based also on the exponentially rising spectrum of hadronic states \cite{Hagedorn:1965st} which predicted the existence of a limiting temperature, then identified as the transition temperature to the QGP.}. The generic name for this new phase of matter is {\it quark--gluon plasma} (QGP). 

The asymptotic state of $T\to\infty$ is a gas of free quarks and gluons, but the situation at experimentally reachable temperatures is not so easy. The temperature at which the transition from a hadron gas to a quark--gluon plasma takes place, $T_c\simeq 160 \div 190$ MeV \footnote{At present there is a theoretical debate on the actual value of this quantity as different lattice groups obtain results approximately in the quoted range. The disagreement seems to be in how the continuum limit is taken, and the corresponding matching from lattice to physical units. See Ref. \cite{Karsch:2008fe,Aoki:2009sc} for details.}, is not large enough to allow the use of  perturbative techniques at temperatures close to it. So, although the use of effective models is still a very active field of research, most of the information about the transition and the properties of the matter in its vicinity comes from lattice QCD calculations, see for example, Ref. \cite{Karsch:2001cy} for a description of the QCD thermodynamics as studied by lattice. 

\begin{figure}
\begin{minipage}{0.5\textwidth}
\begin{center}
\includegraphics[width=0.85\textwidth]{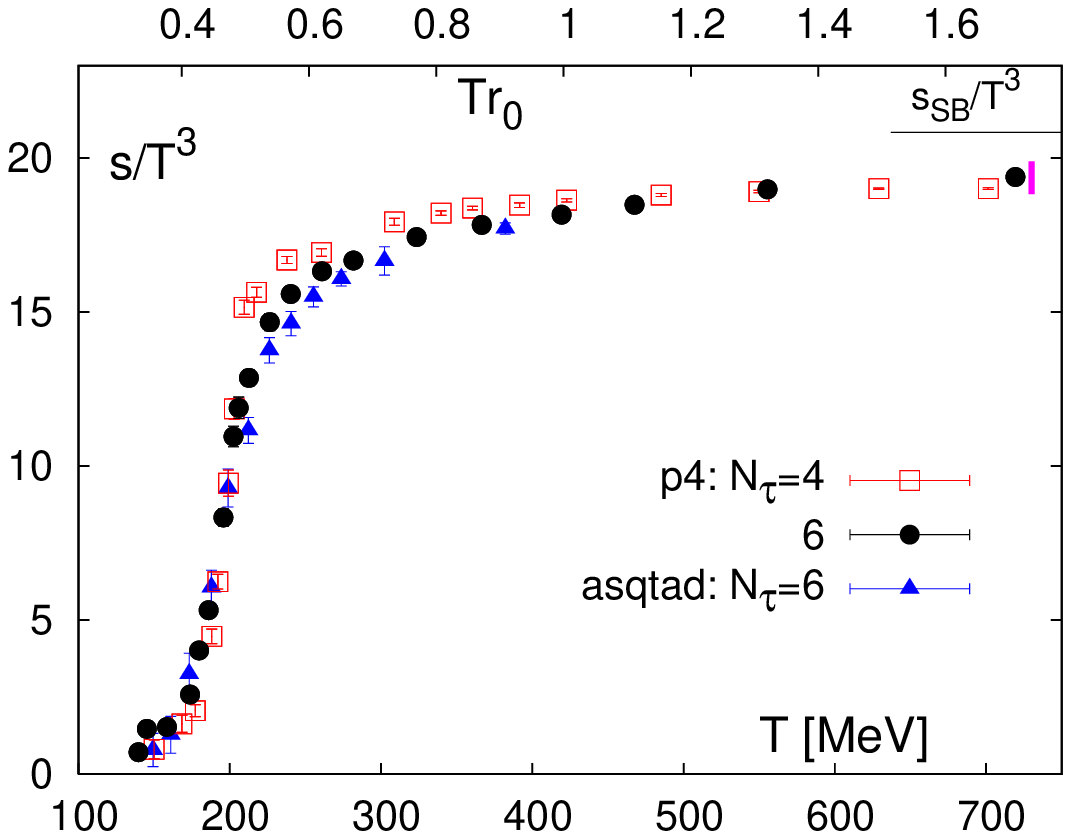}
\end{center}
\end{minipage}
\hfill
\begin{minipage}{0.5\textwidth}
\begin{center}
\includegraphics[width=0.85\textwidth]{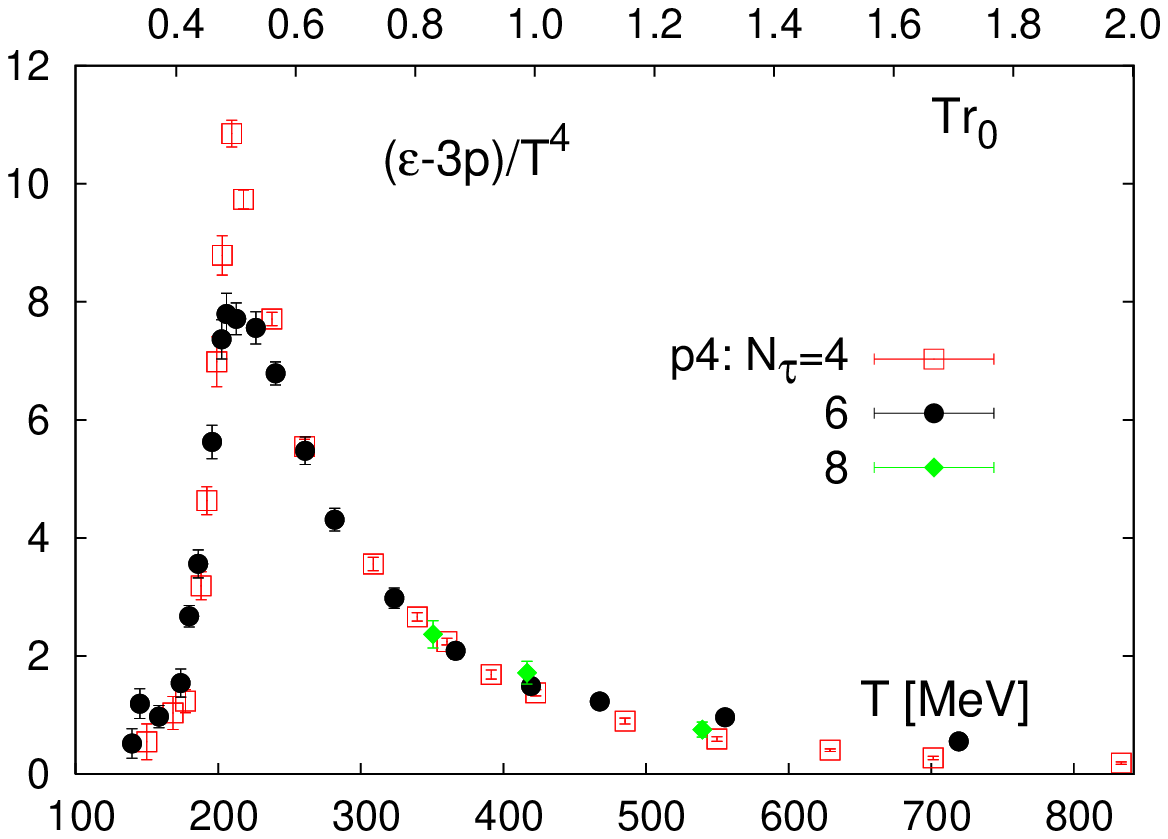}
\end{center}
\end{minipage}
\caption{Left: Entropy, $s\equiv \varepsilon+p$ in units of $T^3$ versus temperature computed in lattice. Right: Trace anomaly in units of $T^4$ as a function of the temperature. Figures from Ref. \protect\cite{Cheng:2007jq}}
\label{fig:EoS}
\end{figure}

A first example of this collective behaviour is the equation of state. The pressure or the energy density measured in units of $T^4$ of an ideal gas is, according to the Stefan--Boltzmann law, proportional to the number of degrees of freedom in the system: so, for a free gas of pions this quantity is $N^\pi_{\rm dof}=3$, while for a free gas of quarks and gluons this quantity is much larger, $N^q_{\rm dof}=2\times 2\times 3$, $N^g_{\rm dof}=2\times 8$, counting spin, colour and (two) flavour states. This different behaviour is observed in lattice calculations where a jump at the transition temperature, $T_c$, appears (Fig. \ref{fig:EoS}). Interestingly, the lattice results also signal a significant departure of the ideal gas behaviour, $\varepsilon=3p$, at temperatures close to $T_c$ (Fig. \ref{fig:EoS}).

In order to understand the nature of the transition order parameters are needed. In QCD with massless quarks, the chiral condensate is the order parameter (see Ref. \cite{Aoki:2006we} for a recent study), while in the infinite mass limit the order parameter is the Polyakov Loop. The order of the transition depend on the actual value of the quark masses and most calculations agree in the absence of a real phase transition at zero baryochemical potential --- the transition would just be a rapid cross-over. 

The behaviour of the potential between heavy quarks is also of interest for phenomenological applications. Although some discussion about the precise meaning and the definition of a potential exists in the case of a hot medium \cite{Mocsy:2008eg}, a screening leading to a non-confining potential is expected to appear at some point above the phase transition and, correspondingly, the heavy quark bound states would cease to exist if this temperature were reached.

\section{General properties of heavy-ion collisions}

Nuclei are very extended objects for all the scales of interest in high-energy physics. For this reason, the geometry of the collision plays a central role in the analysis and interpretation of the experimental results. In the centre-of-mass frame, owing to Lorentz contraction in the longitudinal direction, the two nuclei can be seen as two thin disks of transverse size $2R_A\simeq 2 A^{1/3}$ fm. Some relevant quantities are

\begin{enumerate}

\item The impact parameter which is the distance between the centres of the two colliding nuclei. The impact parameter characterizes the {\it centrality} of the collision. A central collision is one with small impact parameter in which the two nuclei collide almost head-on; a peripheral collision, on the contrary, is one with large impact parameter. A special case is that of ultraperipheral collisions in which the two nuclei interact only through the photons created by the large electromagnetic field of the ions \cite{Baltz:2007kq}.

\item The number of participant nucleons, $N_{\rm part}$, within the colliding nuclei which is the total number of protons and neutrons which take part in the collision. The rest are sometimes called spectators and continue travelling almost unaffected; they are normally measured in forward detectors such as the {\it Zero-Degree Calorimeter} (ZDC) to help in fixing the centrality of the collision.

\item The number of collisions, $N_{\rm coll}$, is the total number of incoherent nucleon--nucleon collisions.

\end{enumerate}

\subsection{The Glauber model and the geometry of the collision}
\label{sec:glauber}

\begin{figure}[hbt]
\begin{center}
\includegraphics[scale=0.4]{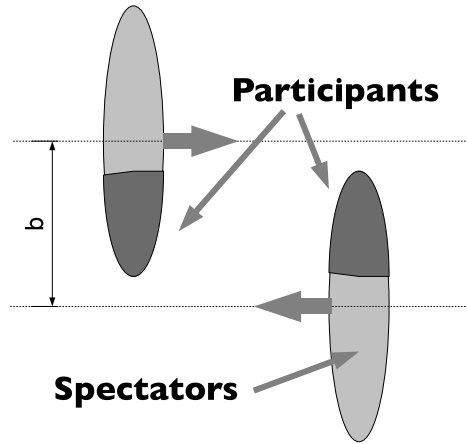}
\end{center}
\caption{Picture for the geometry of a high-energy heavy-ion collision. }
\label{fig:glauberbw}
\end{figure}

The usual way of computing the geometrical quantities enumerated above is by a probabilistic model due to Glauber \cite{glauber}. The starting point is the density of nucleons in a nucleus $A$, $\rho(z,{\bf b})$, for a given longitudinal $z$ and transverse ${\bf b}$ positions. The nuclear profile function or thickness function is the only relevant quantity in most of the calculations and is defined as
\begin{equation}
T_A({\bf b})=\int_{-\infty}^{\infty}dz\rho(z,{\bf b})\, .
\label{eq:prof}
\end{equation}
With the normalization $\int d{\bf b} T_A({\bf b})=1$, the individual probability of a nucleon--nucleon interaction at a given impact parameter is $T_A({\bf b})\sigma_{NN}^{\rm inel}$. If we now consider a proton--nucleus collision, the probability that the proton interacts with $n$ nucleons inside the nucleus is simply given by
\begin{equation}
P(n,{\bf b})=\begin{pmatrix}A\\n\end{pmatrix}\left[1-T_A({\bf b})\sigma_{NN}^{\rm inel}\right]^{A-n} \left[T_A({\bf b})\sigma_{NN}^{\rm inel}\right]^n\, ,
\label{eq:probglauber}
\end{equation}
from which we can compute the number of collisions for a given impact parameter as
\begin{equation}
N_{\rm coll}^{pA}({\bf b})\equiv N_{\rm part}^A({\bf b})=\sum_{n=0}^A nP(n,{\bf b})=AT_A({\bf b})\sigma_{NN}^{\rm inel}\, ,
\label{eq:ncoll}
\end{equation}
where the number of participants of the nucleus $A$, $N_{\rm part}^A$, coincides in this case with the number of collisions $N_{\rm coll}$ --- each nucleon in nucleus $A$ interacts only once. The total number of participants including the colliding proton is then $N_{\rm part}^A+1$. The inelastic cross-section can be computed as
\begin{equation}
\sigma^{\rm inel}_{pA}=\int d{\bf b}\sum_{n=1}^A P(n,{\bf b})=\int d{\bf b}\left[1-\left(1-T_A({\bf b})\sigma_{NN}^{\rm inel}\right)^A\right]\simeq\int d{\bf b}\left[1-\exp\left[-AT_A({\bf b})\sigma_{NN}^{\rm inel}\right]\right]\, ,
\label{eq:painel}
\end{equation}
where the limit of large $A$ has been taken for the last equality. For a nucleus--nucleus collision, the individual probability of nucleon--nucleon interaction at impact parameter ${\bf b}$ is within the {\it optical approximation} --- see for example\cite{d'Enterria:2003qs,Alver:2008aq} for experimental applications.
\begin{equation}
\int d{\bf s}\,T_A({\bf b})T_B({\bf b-s})\sigma_{NN}^{\rm inel}\equiv T_{AB}({\bf b})\sigma_{NN}^{\rm inel}\, .
\end{equation}
$T_{AB}({\bf b})$ is also known as the nuclear overlap function. The corresponding probability of $n$
$NN$ collisions is
\begin{equation}
P(n,{\bf b})=\begin{pmatrix}AB\\n\end{pmatrix}\left[1-T_{AB}({\bf b})\sigma_{NN}^{\rm inel}\right]^{AB-n} \left[T_{AB}({\bf b})\sigma_{NN}^{\rm inel}\right]^n
\label{eq:probglauberNN}
\end{equation}
which gives the inelastic cross-section and number of collisions as
\begin{eqnarray}
\sigma^{\rm inel}_{pA}({\bf b})&=&\int d{\bf b}\left[1-\left(1-T_{AB}({\bf b})\sigma_{NN}^{\rm inel}\right)^{AB}\right]\simeq\int d{\bf b}\left[1-\exp\left[-ABT_{AB}({\bf b})\sigma_{NN}^{\rm inel}\right]\right]\, ,\\
N_{\rm coll}^{AB}({\bf b})&=&\sum_{n=0}^A nP(n,{\bf b})=ABT_{AB}({\bf b})\sigma_{NN}^{\rm inel}\, .
\end{eqnarray}
The  number of participants of nucleus $A$ for a given impact parameter is given by the generalization of (\ref{eq:ncoll}); for this we have to single out one nucleon in nucleus $B$ at transverse position ${\bf b}$ by writing $BT_B({\bf b})$ 
\begin{equation}
N_{\rm part}^A({\bf b})=\int d{\bf s}\, B\,T_B({\bf s})\sigma^{\rm inel}_{pA}({\bf b-s})=\int d{\bf s}\, B\,T_B({\bf s})\exp\left[-AT_A({\bf b-s})\sigma_{NN}^{\rm inel}\right]\, .
\end{equation}
With a similar expression for the number of participants in nucleus $B$, to give the total number of participants as
\begin{equation}
N_{\rm part}({\bf b})=N_{\rm part}^A({\bf b})+N_{\rm part}^B({\bf b})\, .
\end{equation}
In Fig. \ref{fig:glauberbw} a picture of the geometry described above can be found.
%
%
The measure of the geometry in heavy-ion collisions is performed using several methods, with he possibility to cross-check. For example, knowing the number of particles produced in each elementary NN collision, the multiplicity distributions can be related to the number of participants or collisions. Another method is by measuring with ZDC the number of neutrons which did not interact (spectators). In this case, the number of participants is estimated by the total number of nucleons minus the result from the ZDC. As the impact parameter is not a quantity of direct experimental access, a usual way of quoting the centrality of a class of measurements is by a percentage, e.g. 0-5 \% refers to the 5\% most central collisions; 0--50\% to the 50\% most central collisions; 10--20\% to the 20\% most central collisions excluding the 10\% most central ones and so on.

\subsection{The multiplicities}

The total number of particles measured in the detectors after a heavy-ion collision (multiplicity) gives information about the energy density reached. It also gives information about the centrality and, in general, about the main global properties of the created medium. The knowledge of this quantity and its corresponding theoretical description is, then, one main issue in heavy-ion collisions. 

In hadronic collisions, most of the particles are produced at small transverse momentum, belonging to the {\it soft sector} of QCD in which the use of perturbative techniques is not possible. This makes the theoretical description difficult and in most of the cases models are employed. Specially interesting is the formalism based on saturation of partonic densities in the initial nucleus wave function which allows weak coupling calculations to be made in a regime of large occupation numbers. This formalism is generically known as the {\it Colour Glass Condensate}.

\begin{figure}[hbt]
\begin{center}
\includegraphics[scale=0.7]{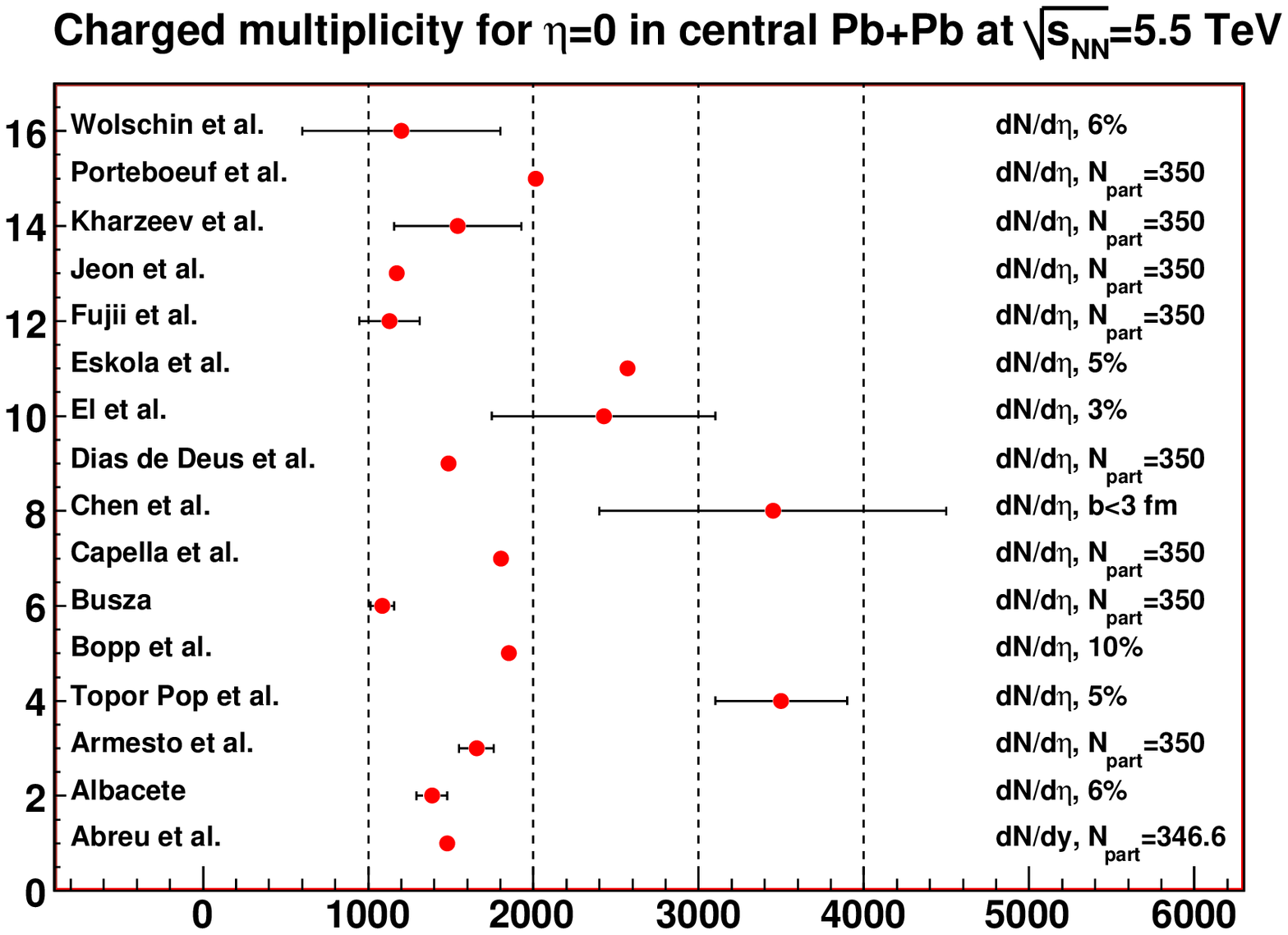}
\end{center}
\caption{Different predictions for the charged particle multiplicities at central rapidities at the LHC. Figure from Ref.\cite{Armesto:2008fj}; results are a compilation from Ref. \cite{Abreu:2007kv}. }
\label{fig:prednestor}
\end{figure}

The most naive estimate of the multiplicity in $AA$ collisions would be that of the corresponding proton--proton collision at the same energy and multiplied by the number of incoherent $NN$ collisions. This number, however, already grossly overestimates the number of produced particles at RHIC, demonstrating the presence of collective behaviour. All the models of multiparticle production prior to RHIC data \cite{Armesto:2000xh} that did not include any type of collectivity failed to reproduce the data. The surviving models present, however, a quite large uncertainty in the extrapolation to the LHC,  see Fig. \ref{fig:prednestor}. In the rest of the section we present some results based on saturation physics. The interested reader can find the relevant references to the different models in the workshop  \cite{Abreu:2007kv}.

\subsubsection{Ideas based on saturation of partonic densities}

\label{sec:satur}

\begin{figure}[hbt]
\begin{center}
\includegraphics[scale=0.5]{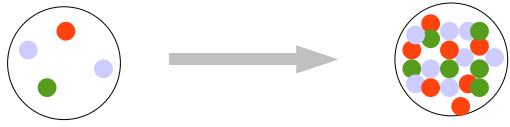}
\end{center}
\caption{Saturation of partonic densities: an intuitive picture of saturated gluon distributions in the transverse plane. Each gluon is depicted by a small disk whose number grows with atomic number and/or energy of the collision.}
\label{fig:satur}
\end{figure}

The density of gluons\footnote{The number of gluons is not a well defined quantity in a field theory like QCD. All the definitions in this section can be done in a more strict way, e.g. by using scattering amplitudes, but this is a useful intuitive picture which contains most of the physics involved.} grows very fast at small $x$. The intuitive picture of saturation of partonic densities is that of overlapping partons, mainly gluons, in transverse plane due to this growth (Fig. \ref{fig:satur}). In the infinite momentum frame,  in which the hadron/nuclei is moving asymptotically fast, the colliding object is a Lorentz-contracted disk of radius $R_A\sim A^{1/3}$. If we do DIS with this object, a virtual photon with virtuality $Q\gg \Lambda_{\rm QCD}$ will scatter incoherently with the partons inside it: {\it counting} the number of partons with transverse size $\sim 1/Q$. Owing to parton multiplication the number of gluons in the proton/nucleus grows very fast with decreasing $x$, i.e., with increasing energy. At some point, the transverse size occupied by the partons is of the same order as the total transverse size of the hadron. This overlap makes gluon fusion probable, with probability ${\cal O}(\alpha_s)$, and the picture of incoherent scatterings breaks down. This defines the saturation scale which can be geometrically estimated as 
\begin{equation}
{\alpha_s}\frac{1}{Q^2_{\rm sat}}A\, xg(x,Q^2_{\rm sat})\simeq \pi R_A^2\, .
\label{eq:qsatest}
\end{equation}
For a gluon density which grows as $xg(x,Q^2)\sim x^{-\lambda}$, the behaviour of the saturation scale as given by this naive geometrical estimate is
\begin{equation}
Q_{\rm sat}^2\simeq x^{-\lambda}A^{1/3}\, .
\label{eq:satest}
\end{equation}
Here, the gluon distribution of bound nucleons is taken as $A\,xg(x,Q^2)$ --- there is no shadowing see below --- and the nuclear radius $R_A\simeq A^{1/3}$. 
%

Intuitively, when $Q_{\rm sat}\gg\Lambda_{\rm QCD}$ this should be the relevant scale in the problem and the parton distributions should depend only on this scale. Indeed, one of the predictions from the Colour Glass Condensate is the scaling of the parton distribution functions with the saturation scale $Q_{\rm sat}^2$ ({\it geometric scaling}). This scale encodes the geometry of the colliding object as well as the dynamics of the gluon multiplication and fusion. 

These ideas lead to a very developed formalism aimed at describing, not only the hadron/nucleus wave function, but also other aspects of QCD at asymptotic energies like multiparticle production or thermalization in heavy-ion collisions, see, for example Refs. \cite{McLerran:2003yx,Mueller:2001fv,Kovner:2005pe,Iancu:2003xm,Gelis:2006dv} for lectures on the topic. Here we present for illustration a simple implementation of this physics based solely on experimental data and the assumption of geometric scaling \cite{Armesto:2004ud}.

{\it i) Geometric scaling in lepton--proton and lepton--nucleus collisions}

All data for the photoabsorption cross-section
$\sigma^{\gamma^* p}(x, Q^2)$ in lepton--proton scattering 
with $x\leq 0.01$ have been found~\cite{Stasto:2000er} to lie 
on a single curve when plotted against the variable $Q^2/\qs^2$, 
with $\qs^2\sim x^{-\lambda}$ and $\lambda\simeq 0.3$. 
Geometric scaling is usually motivated in the QCD dipole 
model~\cite{dipole} where the total $\gamma^*h$ cross-section
reads
\begin{eqnarray}
  \sigma_{T,L}^{\gamma^* h}(x,Q^2) = \int d{\bf r} \hspace{-0.1cm}
  \int_0^1
  \hspace{-0.2cm} dz
  \vert \Psi_{T,L}^{\gamma^*}(Q^2,{\bf r},z)\vert^2\,
  \sigma_{\rm dip}^h({\bf r},x)\, .
  \label{eq1}
\end{eqnarray}
Here $\Psi_{T,L}$ are the perturbatively computed transverse and 
longitudinal wave functions for the splitting of $\gamma^*$ into 
a $q\bar{q}$ dipole of transverse size ${\bf r}$ with light-cone 
fractions $z$ and $(1-z)$ carried by the quark and antiquark respectively.
Both for a proton [$h=p$] and for a nucleus [$h=A$], 
$\sigma_{\rm dip}^h({\bf r},x)$ can be written as an integral of 
the dipole scattering amplitude $N_h$ over the impact parameter 
${\bf b}$,
\begin{eqnarray}
  \sigma_{\rm dip}^h({\bf r},x) = 2 \int d{\bf b}\,
                                     N_h({\bf r},x;{\bf b})\, .
  \label{eq2}
\end{eqnarray}
In this setting, geometric scaling corresponds to the
condition $N_h({\bf r},x;{\bf b})\equiv N_h(r\qsh(x,{\bf b}))$. 
This can be seen by rescaling the impact parameter in (\ref{eq2})
in terms of the radius $R_h$ of the hadronic target, 
${\bf \bar b}={\bf b}/\sqrt{\pi R_h^2}$,
\begin{equation}
  \sigma_{T,L}^{\gamma^* h}(x,Q^2) =  \pi R_h^2
 \int d{\bf r} \hspace{-0.1cm}
  \int_0^1
  \hspace{-0.2cm} dz
  \vert \Psi_{T,L}^{\gamma^*}(Q^2,{\bf r},z)\vert^2 2 \int d{\bf \bar b}\, N_h(r\qsh(x,{\bf \bar b}))\, .
\label{eqnuclsc}
\end{equation}
For a trivial impact-parameter dependence of the saturation
scale, $\qsh(x,{\bf b}) = \qsh(x)\, \Theta(R_h - b)$, and since
$\vert \Psi_{T,L}^{\gamma^*}(Q^2,{\bf r},z)\vert^2$ is
proportional to $Q^2$ times a function of ${\bf r}^2 Q^2$, 
Eq. (\ref{eqnuclsc}) depends solely on $\tau_h = Q^2 / \qsh^2(x)$.
For realistic functional shapes of the form $\qsh(x,{\bf b}) 
\propto f(b/R_h)$, the same $\tau$-dependence results   
if $\qsh^2(x)$ is defined as an appropriate ${\bf b}$-average 
of $\qsh^2(x,{\bf b})$. In the case of $\gamma^* A$ interactions, 
geometric scaling is the property that the $A$-dependence of the
ratio $\sigma_{T,L}^{\gamma^* A}/\pi R_A^2$ can be absorbed in the
$A$-dependence of this impact-parameter independent saturation scale
$\qsa(x)$,
\begin{equation}
  \frac{\sigma^{\gamma^*A}(\tau_A)}{\pi R_A^2}=
  \frac{\sigma^{\gamma^*p}(\tau_A)}{\pi R_p^2}\, .
  \label{eqnormal}
\end{equation}
Now we need to fix the $A$-dependence of the saturation scale, for which we make the assumption --- see (\ref{eq:satest})
\begin{equation}
   \qsa^2=\qs^2\left(\frac{A \pi R_p^2}
   { \pi R_A^2}\right)^\frac{1}{\delta} \hspace{-0.1cm}
   \Rightarrow 
   \tau_A=\tau_p\left(\frac{ \pi R_A^2}{A 
                       \pi R_p^2}\right)^\frac{1}{\delta} ,
   \label{eqtaua}
\end{equation}
where the nuclear radius is given by the usual parametrization
$R_A=(1.12 A^{1/3}-0.86 A^{-1/3})$ fm. We treat $\delta$ and 
$\pi R_p^2$ as free parameters to be fixed by data. 
 
\begin{figure}[hbt]
\begin{center}
\includegraphics[scale=0.5]{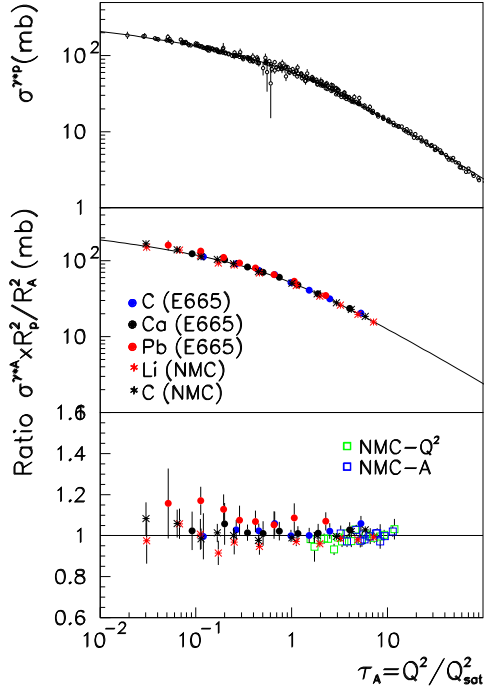}
\end{center}
\caption{Geometric scaling for $\gamma^*p$ (upper panel, data
from Ref.~\cite{proton}),
$\gamma^*A$ (middle panel, data from Refs. \cite{Adams:1995is,Arneodo:1995cs}) 
and the ratio of data for $\gamma^*A$ over the prediction
from (\ref{eqscalf}) (lower panel). As an additional check,
the lower plot also shows data for $\gamma^*A$  
normalized with respect to $\gamma^*C$ \cite{Arneodo:1996rv}, and divided by
the corresponding prediction from Eq. (\ref{eqscalf}).}
\label{figprot}
\end{figure}

In Fig.~\ref{figprot} we plot the experimental $\gamma^*p$ 
data~\cite{proton} with $x\leq 0.01$ as a function of 
$\tau_p= Q^2 / \qs^2$. For $\qs^2$, we use in this plot
the  Golec-Biernat and W\"usthoff (GBW) 
parametrization~\cite{Golec-Biernat:1998js} 
with $\qs^2=(\bar x/x_0)^{-\lambda}$ in GeV$^2$,
$x_0= 3.04\cdot 10^{-4}$ and $\lambda=0.288$.  
In order to proceed, we need a parametrization of the experimental data (a functional form for the scaling curve). We find that the data~\cite{proton} are well parametrized by 
\begin{eqnarray}
  \sigma^{\gamma^* p}(x,Q^2) \equiv
  \Phi(\tau_p) = 
\bar\sigma_0
  \left[ \gamma_E + \Gamma\left(0,\xi\right) +
         \ln\xi \right]\, ,
       \label{eqscalf}
\end{eqnarray}
where $\gamma_E$ is the Euler constant, $\Gamma\left(0,\xi\right)$
the incomplete $\Gamma$ function, and $\xi=a/\tau_p^b$,
with $a=1.868$ and $b=0.746$. The normalization is fixed by
$\bar\sigma_0=40.56$ mb. 
%

To determine $\qsa^2$, we compare the functional shape of 
(\ref{eqscalf}) to the available experimental data for $\gamma^*A$ 
collisions with $x\leq 0.0175$ 
\cite{Adams:1995is,Arneodo:1995cs,Arneodo:1996rv}, using 
$\xi = a/\tau_A^b$.
The parameters $\delta$ and $\pi R_p^2$ in Eqs.
(\ref{eqnormal})--(\ref{eqscalf}) are fitted by $\chi^2$ 
minimization adding the statistical 
and systematic errors in quadrature. The data sets \cite{Adams:1995is}, 
\cite{Arneodo:1995cs}, and 
\cite{Arneodo:1996rv} have additional normalization errors of 0.4\%, 0.2\%, and 
0.15\%; the quality of the fit improves by multiplying the data
by the factors 1.004, 1.002, and 0.9985, respectively. We
obtain $\delta=0.79\pm0.02$ and $\pi R_p^2=1.55 \pm 0.02$
fm$^2$ for a $\chi^2/{\rm dof} = 0.95$ --- see Fig.~\ref{figprot} for 
comparison. If the normalizations are all set to 1,
we obtain an almost identical fit with $\delta=0.80\pm 0.02$
and $\pi R_p^2=1.57 \pm 0.02$ fm$^2$ for a $\chi^2/{\rm dof} = 1.02$. 
If we impose $\delta=1$ in the fit, which corresponds to $\qsa^2\propto 
A^{1/3}$ for large nuclei, a much worse value of
$\chi^2/{\rm dof} =2.35$ is obtained. We conclude that the small-$x$
experimental data on $\gamma^*A$ collisions favor an increase of 
$\qsa^2$ faster than $A^{1/3}$. The numerical coincidence 
$b\simeq \delta$ is consistent with the absence of shadowing
in nuclear parton distributions at $Q^2\gg \qsa^2$.

{\it ii) Geometric scaling and multiplicities in heavy-ion collisions}

The fact that geometric scaling is found both in lepton-proton and -nucleus data is interesting, but we can extend the formulation a bit further and consider the case of multiplicities in symmetric hadronic or heavy-ion collisions --- this means the same colliding systems and measure at central rapidities. This allows us to make several simplifications as only one saturation scale is involved. For the discussion, consider for simplicity a model of multiparticle production in which a factorization between the gluon distributions $\phi(y,{\bf k},{\bf b})$ of the two nuclei $A$ and $B$ appears~\cite{Gribov:tu} 
\begin{equation}
   \frac{dN^{AB}_g}{dyd{\bf p}_t d{\bf b}}\propto \frac{\alpha_S}{{\bf p}_t^2}
   \int d{\bf k}\  \phi_A(y,{\bf k}^2,{\bf b})\,
    \phi_B\left(y,({\bf k}-{\bf p}_t)^2,{\bf b}\right)\, ,
   \label{eqfact}
\end{equation}
where $\phi_h(y,{\bf k},{\bf b})=\int d{\bf r}\,\exp\{i{\bf r\cdot k}\}\,
N_h({\bf r},x;{\bf b})/(2\pi r^2)$~\cite{facto}, 
$y = \ln 1/x $.
For geometric scaling, $\phi_A(y,{\bf k}^2,{\bf b})$$\equiv$
$\phi({\bf k}^2/\qsa^2(y,{\bf b}))$, we find the dependence 
\begin{eqnarray}
  &&\frac{dN^{AA}_g}{dy}\Bigg\vert_{y\sim 0}\propto 
  \int \frac{d{\bf p}_t}{{\bf p}_t^2} d{\bf k} d{\bf b}\ 
  \phi\left(\frac{{\bf k}^2}{\qsa^2}\right)\,
  \phi\left(\frac{({\bf k}-{\bf p}_t)^2}{\qsa^2}\right)\nonumber \\
  &&=\qsa^2 \pi R_A^2\int \frac{d{\bf s}}{{\bf s}^2} d{\bf \tau} 
  d{\bar {\bf b}}\ 
  \phi({\bf \tau}^2)\,\phi\left( ({\bf \tau}-{\bf s})^2\right).
  \label{eqsym}
\end{eqnarray}
The integral is then independent of the colliding objects --- geometric scaling has been used --- resulting in
\begin{equation}
  \frac{dN^{AA}}{dy}\Bigg\vert_{y\sim 0}\propto 
  \qsa^2\, \pi R^2_A\, .
\label{eqmprop}
\end{equation}
Similar expressions  arise in different models of 
hadroproduction~\cite{Gribov:tu,facto, Kovchegov:2000hz,Eskola:1999fc}.
It is worth noting that factorization is not  actually needed in (\ref{eqfact}), 
any integrand with $(k/\qsa)$--scaling leads
to Eq.~(\ref{eqmprop}), see Refs.~\cite{Eskola:1999fc,Kovchegov:2000hz}.
In all these models, the hadron yield is assumed to be proportional
to the yield of produced partons.

\begin{figure}
\begin{center}
\includegraphics[scale=0.5]{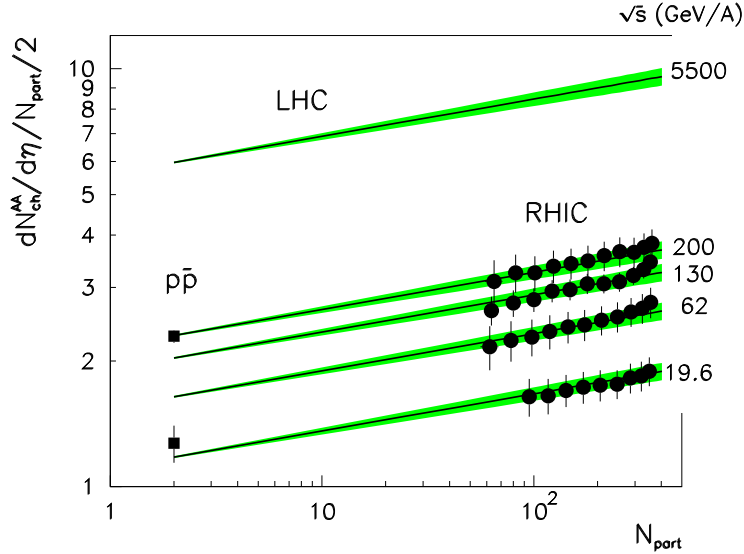}
\end{center}
\caption{Energy and centrality dependence of the multiplicity of charged 
particles in Au+Au collisions (\ref{eqmult}) compared to PHOBOS data
\cite{Back:2002uc}.
Also shown in the lower panel are the ${\bar p}$+$p$ data 
\cite{protmult} and results for $\sqrt{s}=$  5500 GeV/A.}
\label{figmult}
\end{figure}

Equation~(\ref{eqmprop}) relates the energy and nuclear size dependence of the multiplicities to the corresponding quantities measured in lepton-hadron collisions. In particular, the energy dependence is given by the 
GBW parameter $\lambda=0.288$. For the centrality dependence 
of (\ref{eqmprop}), we use the known proportionality in symmetric
A+A collisions between 
the number $N_{\rm part}$ of participant nucleons 
and the nuclear size $A$. With
$\qsa^2 \propto A^{1/3\delta}$, and $\delta=0.79\pm 0.02$,
we thus obtain\footnote{Notice a factor of 2 missing in the original paper \cite{Armesto:2004ud}.},
\begin{equation}
\frac{2}{N_{\rm part}}
\frac{dN^{AA}}{d\eta}\Bigg\vert_{\eta\sim 0}=N_0\sqrt{s}^\lambda 
N_{\rm part}^{\frac{1-\delta}{3\delta}}\, .
\label{eqmult}
\end{equation}
As seen in Fig.~\ref{figmult}, this ansatz accounts for experimental data 
from the PHOBOS Collaboration~\cite{Back:2002uc} on charged multiplicities 
in Au+Au collisions at $\sqrt{s}= 19.6$, 62, 130, and 200 GeV/A. 
Even the ${\bar p}$+$p$ data (\cite{protmult}, as quoted in Ref. \cite{Back:2002uc})
at $\sqrt{s}=$ 19.6 and 200 GeV are accounted for by Eq.~(\ref{eqmult}). 
Since all data are at mid-rapidity, the Jacobian between rapidity $y$ and
pseudorapidity $\eta$ is approximately constant. It has been
absorbed in the overall normalization  $N_0=0.47$ which is
independent of the energy and the centrality of the collision.
Figure~\ref{figmult} also shows the prediction of (\ref{eqmult})  for LHC energy 
($\sqrt{s}=5500$ GeV/A) --- this result is quoted in Fig. \ref{fig:prednestor} as ``Armesto et al'' --- and for smaller colliding nuclei. Equation~(\ref{eqmult}) 
implies that the energy and the centrality dependence of the 
multiplicity factorize, in agreement with the results by 
PHOBOS~\cite{Back:2002uc}.
%

{\it iii) Multiplicities computed from the QCD evolution equations}

\begin{figure}
\begin{center}
\includegraphics[scale=0.4]{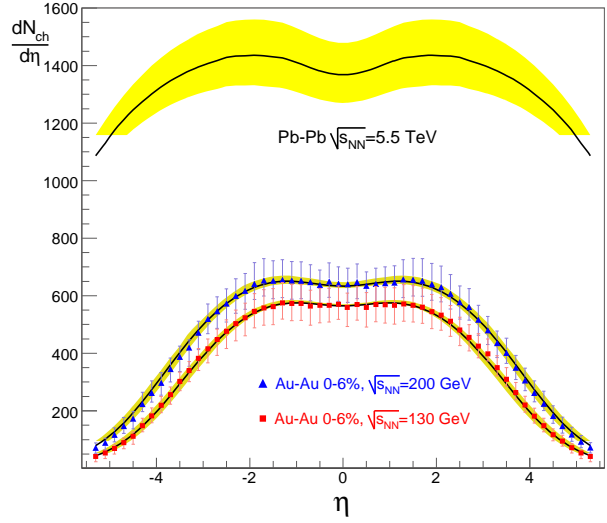}
\end{center}
\caption{Rapidity and energy dependence of the multiplicity of charged 
particles in Au+Au collisions (\ref{eqmult}) compared to PHOBOS data
\cite{Back:2002uc} and the corresponding extrapolation to LHC energies.}
\label{fig:javier}
\end{figure}

Although the CGC approach is a very advanced theoretical construction, its implementation in actual phenomenological analysis is difficult, due, in part to the complexity of the equations to be solved and in part to unknowns in, for example, the initial conditions for evolution equations, etc. For this reason, most of the analyses of the data in the last years made strong simplifications or, simply, tried to find generic features in the data expected from the CGC such as the geometric scaling described above. Another difficulty was the use of leading order results which are known to produce too fast evolution of the parton distributions. A breakthrough in this direction is the computation \cite{Balitsky:2006wa,Kovchegov:2006vj,Albacete:2007yr} of part of the non-leading corrections to the equations, leading to dependences in agreement with expectations from data. These equations are solved numerically in Ref. \cite{Albacete:2007sm} to describe the energy and rapidity dependences of the multiplicities in nuclear collisions for the first time using first-principle calculations of the evolution equations. The results of this study are plotted in Fig. \ref{fig:javier} and quoted in Fig. \ref{fig:prednestor} as ``Albacete''.

\section{Soft probes and the bulk}

The soft part of the produced spectrum provide very useful tools to characterize the collective properties of the nuclear collisions --- and eventually the properties of a new state of matter. It is not our purpose here to give an extensive review of all probes proposed in heavy-ion collisions. We will, instead, limit ourselves to the ones specially relevant in the last years, with the advent of collider energies in nuclear collisions at RHIC. 

A powerful way of looking at the degree of collectivity reached in a heavy-ion collision is by studying the azimuthal distribution of particles produced at a given transverse momentum $p_T$: if the collisions were all independent, the particles would be uniformly distributed (within statistical fluctuations) while the presence of azimuthal anisotropies would indicate that non-trivial phenomena are taking place. The hydrodynamic modeling of the heavy-ion collisions attempts to describe such phenomena.

\subsection{The hydrodynamical evolution}

The use of hydrodynamics to solve the space--time evolution of hadronic collisions goes back to Landau in the 1950s \cite{Landau:1953gs} and has been established as one of the main fields of research in heavy-ion collisions. In those dynamical situations in which the mean free path of the particles in the medium is very small, medium properties such as energy density, pressure, or temperature can be described in a hydrodynamical approach. The basic equation corresponds to the conservation of the energy-momentum tensor
\begin{equation}
\partial_\mu T^{\mu\nu}=0\, .
\label{eq:hydro}
\end{equation}
Neglecting viscosity, the energy-momentum tensor can be written as
\begin{equation}
T^{\mu\nu}=(\epsilon+p)u^\mu u^\nu-pg^{\mu\nu}\, ,
\label{eq:emtensor}
\end{equation}
where $\epsilon$, $p$, and $u^\mu$ are the (local) energy density, pressure, and four-velocity of a fluid element. The latter is normalized as $u^\mu u_\mu=1$. Lectures on hydrodynamical modelling of heavy-ion collisions can be found, for example, in Refs. \cite{Rischke:1998fq,Hirano:2008hy}.

\subsubsection{The Bjorken model}

A very simple model for the hydrodynamical description of a heavy-ion collision due to Bjorken \cite{Bjorken:1982qr} captures some of the properties of the evolution, neglecting all the transverse dynamics: Assume the collision of two big nuclei at very high energy so that, owing to Lorentz contraction, they can be considered as two disks of infinite transverse size. Define the spatial rapidity and the proper times as
\begin{eqnarray}
y=\frac{1}{2}\log\frac{t+z}{t-z} & ; & \tau=\sqrt{t^2+z^2}\, .
\label{eq:raptau}
\end{eqnarray}
At asymptotic collision energies, boost invariance is a good approximation if we are interested in the central rapidity region. To see this, recall that a boost corresponds to an additive term in the rapidity, $y'=y+y_{\rm boost}$; if the energy of the colliding objects is very large, small  longitudinal boosts (compared with the collision energy) should not affect the results, i.e., the result does not depend on the rapidity. In this case, the initial conditions are
\begin{eqnarray}
p(\tau)\,; & \epsilon(\tau) \, ; & u^\mu=\gamma(1,0,0,z/t)\, ,
\end{eqnarray}
and the hydrodynamic equations simplify to
\begin{equation}
\frac{d\epsilon}{d\tau}+\frac{\epsilon+p}{\tau}=0\, .
\label{eq:bjorken}
\end{equation}
We need an equation of state, relating $\epsilon$ and $p$, to close the equations. For an ideal EoS, $\epsilon=3p$, the solutions are
\begin{eqnarray}
\epsilon(\tau)=\frac{\epsilon_0}{\tau^{4/3}}\, ; & T(\tau)=\frac{T_0}{\tau^{1/3}}\, ,
\label{eq:bjorkensol}
\end{eqnarray}
which tells how the energy density or the temperature decrease with proper time due to longitudinal flow in a boost-invariant ideal fluid.

\subsection{Transverse flow produced by hydrodynamics}

\begin{figure}
\begin{center}
\includegraphics[scale=0.6]{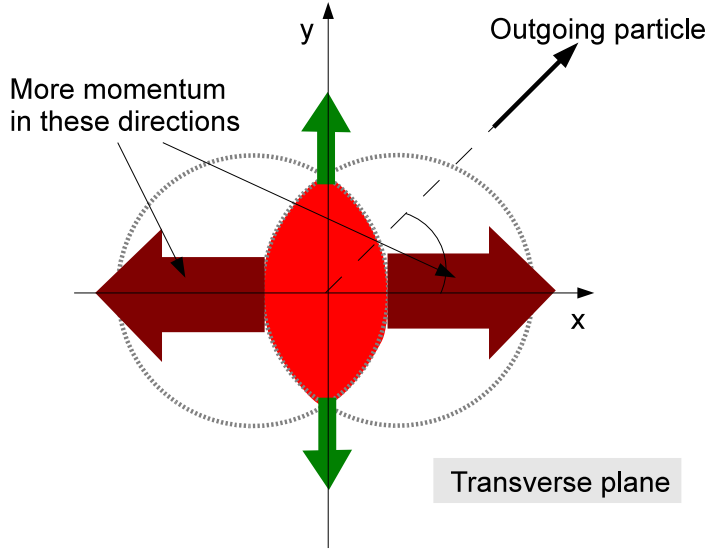}
\end{center}
\caption{Hydrodynamics of a non-central nucleus--nucleus collision in the transverse plane.}
\label{fig:elliptic}
\end{figure}

The above model is a very simplified version of the hydrodynamical evolution and, in particular, one very important feature of realistic heavy-ion collisions, the presence of transverse expansion, is not described. The sophisticated hydrodynamical calculations used nowadays  include all these features, but for our purpose it is instructive to consider the Euler equation, contained in Eq. (\ref{eq:hydro})
\begin{equation}
\frac{d\beta}{dt}=-\frac{c^2}{\epsilon+p}\nabla p\, .
\label{eq:euler}
\end{equation}
Equation (\ref{eq:euler}) tells us that gradients of pressure produce acceleration of fluid elements. This is especially relevant for the transverse plane, where, taking again the ideal case $p=\epsilon/3$, gradients of energy density are defined by the geometry of the collision. These transverse energy density profiles translate into a transverse expansion of the medium. The most interesting case is that of non-central collisions in which the medium has a characteristic {\it almond shape} with a strong azimuthal asymmetry in the energy density profile. This spatial asymmetry leads, hence, to an azimuthal-dependent transverse flow, i.e., the momentum of the particles after the freeze-out is strongly dependent on the azimuthal angle. In Fig. \ref{fig:elliptic} an intuitive picture of the effect is depicted: i) two nuclei collide semiperipherally and the overlapping area is non-symmetric in the transverse plane; ii) the gradient of energy density is larger in the direction labelled as $x$ than in the one labelled as $y$ --- these directions are normally called {\it in-plane} and {\it out-of-plane} for a plane, {\it reaction plane}, defined by the two beam directions; iii) this asymmetry leads to different density and pressure fields, producing azimuthally dependent accelerations, maximal in the $y$ direction and minimal in the $x$ direction.

The transverse flow is one of the most important measurements in favor of the hydrodynamical behaviour in heavy-ion collisions and it is normally parametrized by the first non-trivial coefficient (at mid-rapidity) of the Fourier expansion in the azimuthal angle, called $v_2$,
\begin{equation}
\frac{dN}{d\phi}\propto 1+2\,v_2\,\cos(2\phi)\, .
\label{eq:v2}
\end{equation}
A value of $v_2\ne 0$ signals the presence of azimuthal asymmetries in the final state. Note that for a finite number of particles ($N$) these asymmetries are always present, in an event-by-event study, due to limited statistics; in the extreme case of two particles in the final state the asymmetry is maximized as energy-momentum conservation implies they are back-to-back. These statistical asymmetries are, however, reduced as $1/\sqrt{N}$. 

\begin{figure}
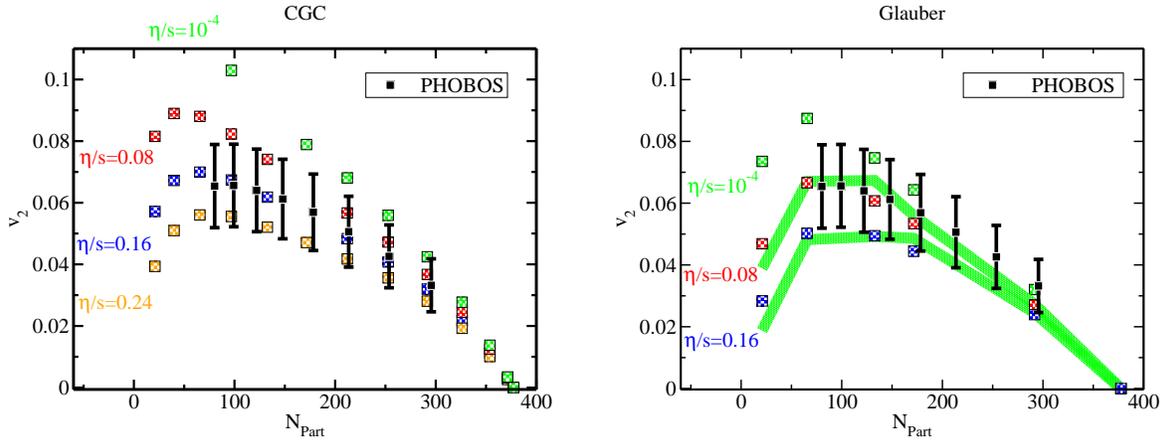

\begin{minipage}{0.5\textwidth}
\begin{center}
\includegraphics[scale=0.3]{high-quality-v2int-momecc2.eps}
\end{center}
\end{minipage}
\begin{minipage}{0.5\textwidth}
\begin{center}
\includegraphics[scale=0.3]{high-quality-v2intnew.eps}
\end{center}
\end{minipage}
\caption{Comparison of a hydrodynamical calculation~\cite{Luzum:2008cw} with experimental data from \cite{Alver:2007sp}. The effects of different values for the viscosity and the initial conditions in the hydrodynamical evolution are studied.}
\label{fig:v2}
\end{figure}

At present the use of hydrodynamical models is one of the most active subject areas in heavy-ion collisions. State-of-the-art calculations are beginning to include viscous corrections to the numerical simulations. In Fig. \ref{fig:v2} we present the comparison with data of one from these studies \cite{Luzum:2008cw}.

\section{Hard Probes}

The study of processes with large virtualities involved, such as production of particles with large masses and/or large transverse momentum, has become one of the main fields of research in heavy-ion collisions. In these hard processes different scales appear and can be separated into the different contributions to the factorized cross-section, which can be symbolically written as
\begin{equation}
\sigma^{AB\to h}=
f_A(x_1,Q^2)\otimes f_B(x_2,Q^2)\otimes \sigma(x_1,x_2,Q^2)\otimes D_{i\to h}
(z,Q^2)\, .
\label{eq:hard}
\end{equation}
Here, the large virtuality, $Q^2$, allows us to compute the partonic cross-section, $\sigma(x_1,x_2,Q^2)$, as a perturbative expansion in powers of $\alpha_s(Q^2)$. This perturbative partonic process takes place in a short time $\sim 1/Q$, i.e., it describes the processes associated to short-distances. In collisions involving hadrons, there is always a non-perturbative contribution associated to long-distances (small scales ${\cal O}(\Lambda_{\rm QCD})$) such as the size of the hadron. Two of them appear in (\ref{eq:hard}): on the one hand, the parton distribution functions (PDF) $f_A(x,Q^2)$, encoding the partonic structure of the colliding objects at a given fraction of momentum $x$ and virtuality $Q$; and the fragmentation functions (FF), $D(z,Q^2)$, describing the hadronization of the parton $i$ into a final hadron $h$ with a fraction of momentum $z$. In the nuclear case, these are the quantities which are modified when the {\it extension} of the colliding system interferes with the dynamics, while the short-distance part is expected to remain unchanged if the virtuality is large enough. These modifications could involve the non-perturbative initial condition as well as the evolution equations. In the latter case, non-linear terms become important.

At the LHC, where proton--proton, proton--nucleus and nucleus--nucleus collisions will occur, a good knowledge of the PDFs is essential: proton PDFs are needed to compute the cross-sections of processes which are background for searches such as the Higgs, SUSY, etc.; nuclear PDFs are also needed for benchmarking those effects in nucleus--nucleus collisions not related with the production of the hot medium under study. 

Although the presence of the non-perturbative PDFs is unavoidable at the LHC, there are processes in which the long-distance term {\it after} the hard collision, the FF in Eq. (\ref{eq:hard}), is not present. Some examples are photon, $Z$, or $W^\pm$ production which are produced directly in the hard scattering. These processes, well described by perturbative techniques, also have an important role as benchmarks in heavy-ion collisions.

The employement of hard processes as tools to characterize the medium properties uses the fact that the non-perturbative, long-distance, terms in (\ref{eq:hard}) are affected by the presence of a macroscopic medium. A conceptually simple example is the case of the $J/\Psi$, whose production cross-section can be written as
\begin{equation}
\sigma^{hh\to J/\Psi}=
 f_i(x_1,Q^2)\otimes f_j(x_2,Q^2)\otimes
\sigma^{ij\to [c\bar c]}(x_1,x_2,Q^2)
 \langle {\cal O}([c\bar c]\to J/\Psi)\rangle\, ,
 \label{eq;jpsi}
\end{equation}
where now $ \langle {\cal O}([c\bar c]\to J/\Psi)\rangle$ is a purely non-perturbative quantity  describing the hadronization of a $c\bar c$ pair in a given state (for example a colour octet) into a final $J/\Psi$. In the case where a hot medium is formed in the collision we can expect that the potential between the heavy-quarks will be screened and hence the bound states cannot survive  \cite{Matsui:1986dk}. Ideally this would correspond to a strong reduction of the hadronization term $ \langle {\cal O}([c\bar c]\to J/\Psi)\rangle$ leading to a suppression of the yield of $J/\Psi$'s if the hot medium is created. In practice, however, the theoretical control over the different terms in (\ref{eq:jpsi}) and their nuclear modifications is not good enough and the interpretation of the experimental results is still not clear.

Another well-known experimental example of these modifications is the strong suppression of particles produced at large transverse momentum in heavy-ion collisions \cite{RHIC,SPS}. The effects due to the presence of a hot medium in the structure of the high-$p_T$ part of the produced spectrum is known under the generic name of {\it jet quenching}, see, for example, Refs. \cite{Baier:2000mf,Kovner:2003zj,Gyulassy:2003mc,CasalderreySolana:2007zz}.

In the next sections we report on the present knowledge of some of these hard processes see, for example, \cite{d'Enterria:2006su,Jacobs:2004qv,Muller:2006ee,Hippolyte:2009xz} for additional experimental results.

\subsection{Nuclear parton distribution functions}

It is a well-stablished fact that the partonic structure of high-energy nuclei is different from the incoherent superposition of the component nucleons, see Ref. \cite{Armesto:2006ph} for a recent review. Indeed, experiments of DIS with nuclei have measured different nuclear effects in the whole range of $x$ which can be associated to modifications of the nuclear PDFs when compared with the free proton ones. In particular, it has been customary to give different names to these modifications depending on the relevant range of $x$ under consideration: 1) {\it shadowing} for the suppression observed at small ($x\lesssim 0.05$); 2) {\it antishadowing} for the enhancement at moderate values of $0.05\lesssim x\lesssim 0.3$; 3) {\it EMC effect} for the suppression observed in the region $0.3\lesssim x\lesssim 0.7$; and 4) {\it Fermi motion} for the enhancement when $x\to 1$.

The observed effects on the structure functions measured in DIS with nuclei can be implemented as modifications of the non-perturbative input in a DGLAP analysis of the nuclear PDFs when compared with the free proton case. This type of analyses has reached a degree of sophistication similar to the ones for free protons, with studies done at LO and NLO and error analysis following the Hessian method. However, the small amount of experimental data, especially the region of small-$x$, makes the knowledge of the nPDFs not as precise as will be necessary for a correct interpretation of all the experimental data at the LHC. This deficiency will be cured at the LHC with the presence of a parallel proton--nucleus programme.

\subsubsection{Global fits}

 As in the case of the free proton, the sets of nuclear PDFs are obtained by global fits to different sets of experimental data \cite{Eskola:1998iy,Eskola:1998df,Eskola:2007my,Hirai:2001np,Hirai:2004wq,deFlorian:2003qf,Hirai:2007sx,Schienbein:2007fs,Eskola:2008ca,Eskola:2009uj}. Nowadays, all groups include data on DIS with nuclei and Drell-Yan production in proton--nucleus collisions. Data on inclusive particle production in deuterium-gold collisions measured at RHIC was first included in the analysis of Ref. \cite{Eskola:2008ca}, showing, in particular, the feasibility of these analyses and the uncertainties to be faced when pA data from the LHC becomes available. 
 
Most of the groups perform the analysis at the level of the ratios
\begin{equation}
R^A_i(x,Q^2)\equiv\frac{f_i^A(x,Q^2)}{f_i(x,Q^2)}\, ,
\label{eq:ratiospdfs}
\end{equation}
defined as the ratio of the PDF for a proton inside the nucleus, $f_i^A(x,Q^2)$, to the corresponding one for a free proton, $f_i(x,Q^2)$. Assuming a given functional form for the ratios (\ref{eq:ratiospdfs}) at an initial scale $Q_0^2$, their parameters are obtained by evolving both the numerator and the denominator by the DGLAP evolution equations \cite{DGLAP} and fitting the used sets of data. 

\begin{figure}[!htb]
\center
\includegraphics[scale=0.55]{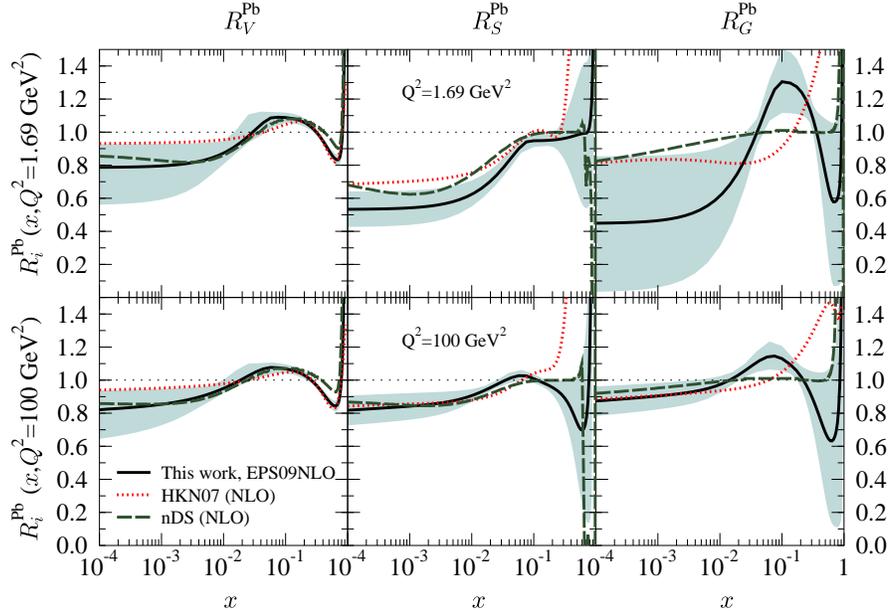}
\caption[]{\small Comparison of the average valence and sea quark, and gluon modifications at $Q^2 = 1.69 \, {\rm GeV}^2$ and $Q^2 = 100 \, {\rm GeV}^2$ for Pb nucleus from the NLO global DGLAP analyses HKN07~\cite{Hirai:2007sx}, nDS~\cite{deFlorian:2003qf}, and EPS09NLO  \cite{Eskola:2009uj} including the error estimates. Figure from Ref. \cite{Eskola:2009uj}.}
\label{fig:pdfscomp}
\end{figure}

The procedure to perform nPDF global fits follows closely that of the free proton case:
\begin{enumerate}
\item Take a functional form for the ratios at the initial scale $Q_0^2\simeq 1$ GeV$^2$ with a number of free parameters $\{z\}$: $R_i(A,x,Q^2_0,\{z\}),\, i=g,\,u,\,\bar u,\,d,\,\bar d,\,s\dots$

\item Assuming a known set of free proton PDFs, evolve the numerator and denominator of (\ref{eq:ratiospdfs}) separately to a larger scale $Q^2$ using the Dokshitzer--Gribov--Lipatov--Altarelli--Parisi (DGLAP) equations \cite{DGLAP}. 

\item Compute the theoretical values corresponding to each data set. This gives a value of $\chi^2$ for a particular choice of the parameters $\{z\}$
\begin{equation}
\chi^2(\{z\})   \equiv  \sum _N \chi^2_N(\{z\}).
\end{equation}
\item If the function defined by the parameters $\{z\}$ is the minimum of $\chi^2$ this is the final answer, otherwise, choose another $\{z\}$ and start again from point 2.
\end{enumerate}

\begin{figure}[htbp]
\begin{minipage}{0.33\textwidth}
\center
\includegraphics[scale=0.27]{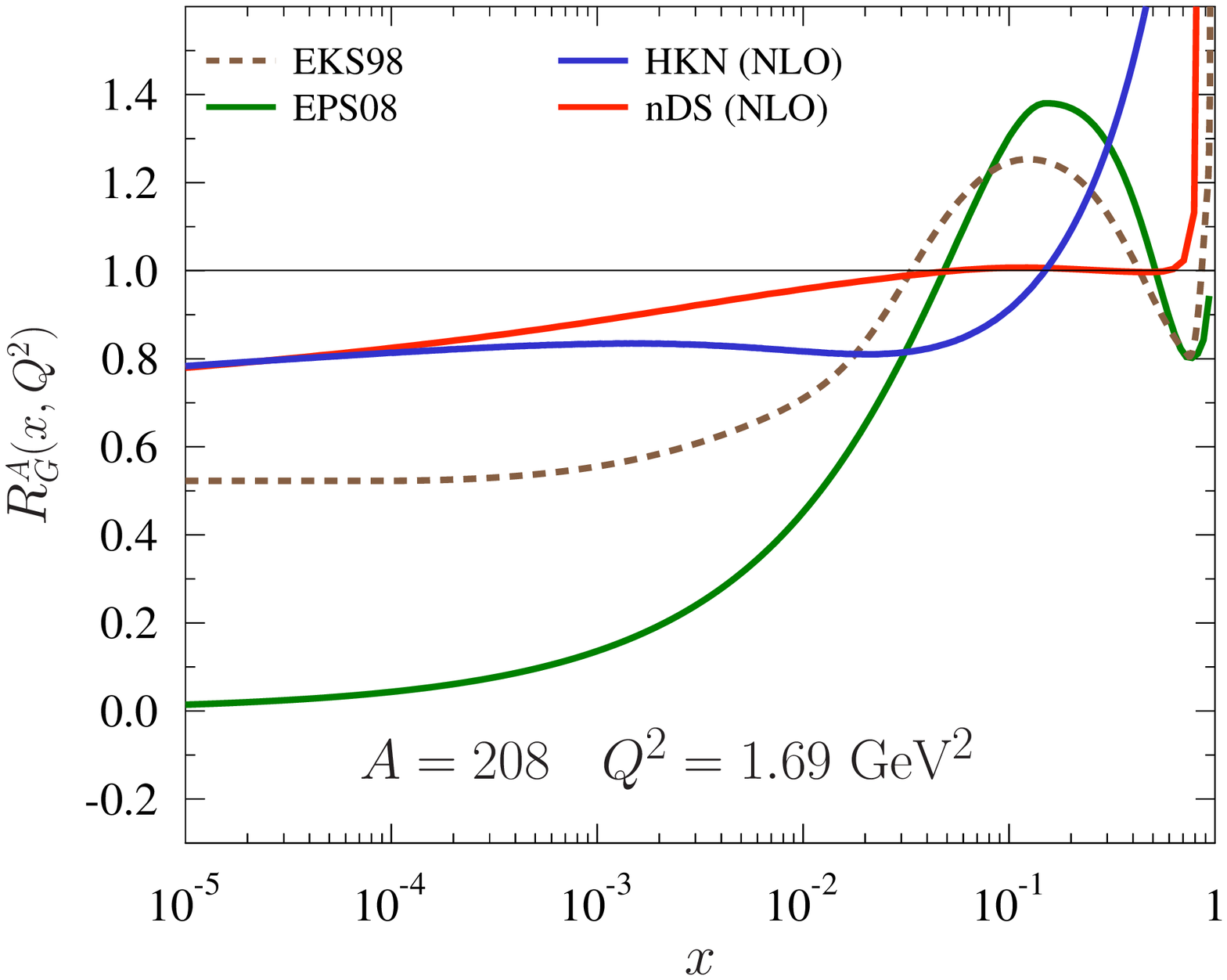}
\end{minipage}
\begin{minipage}{0.33\textwidth}
\center
\includegraphics[scale=0.27]{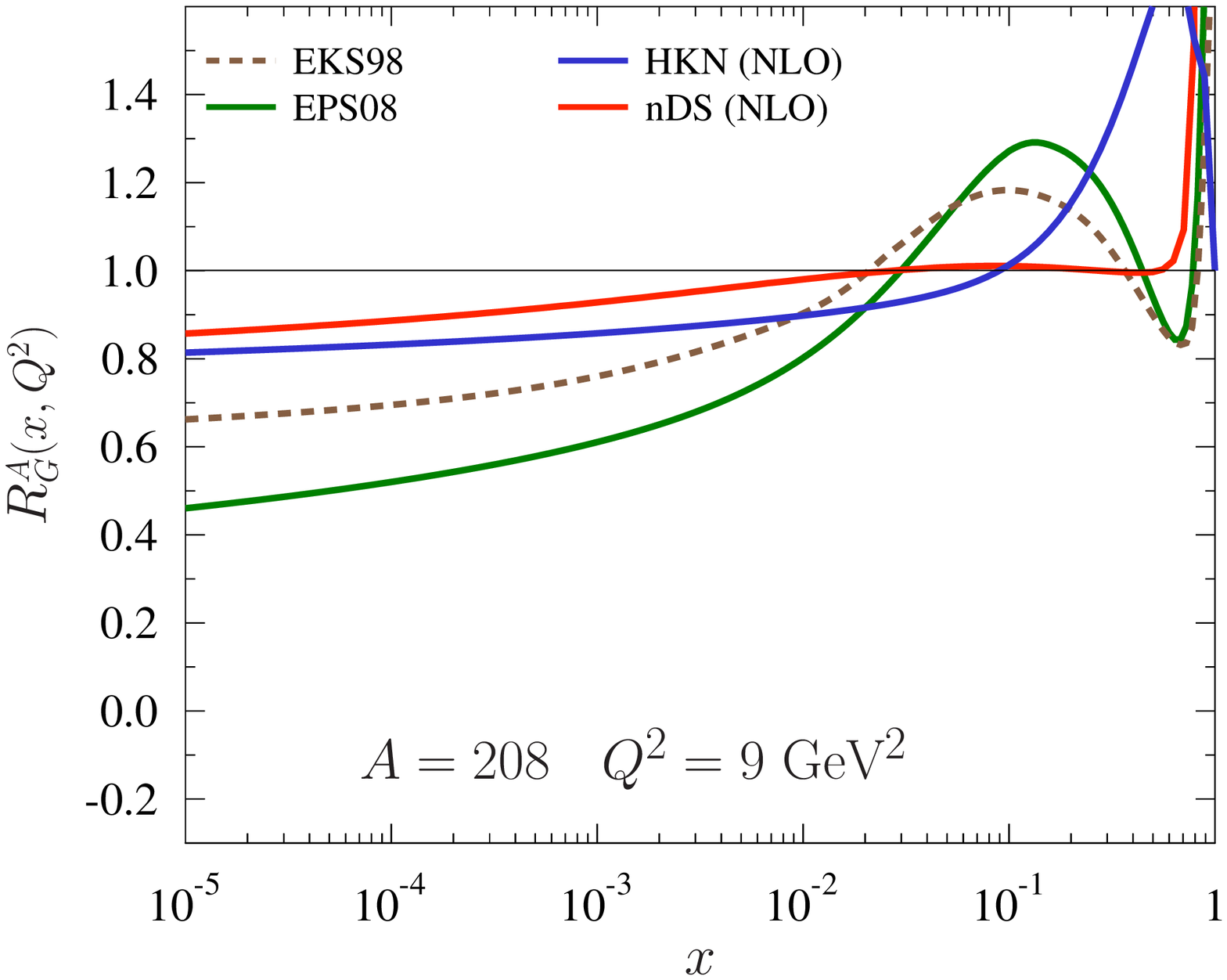}
\end{minipage}
\begin{minipage}{0.33\textwidth}
\center
\includegraphics[scale=0.27]{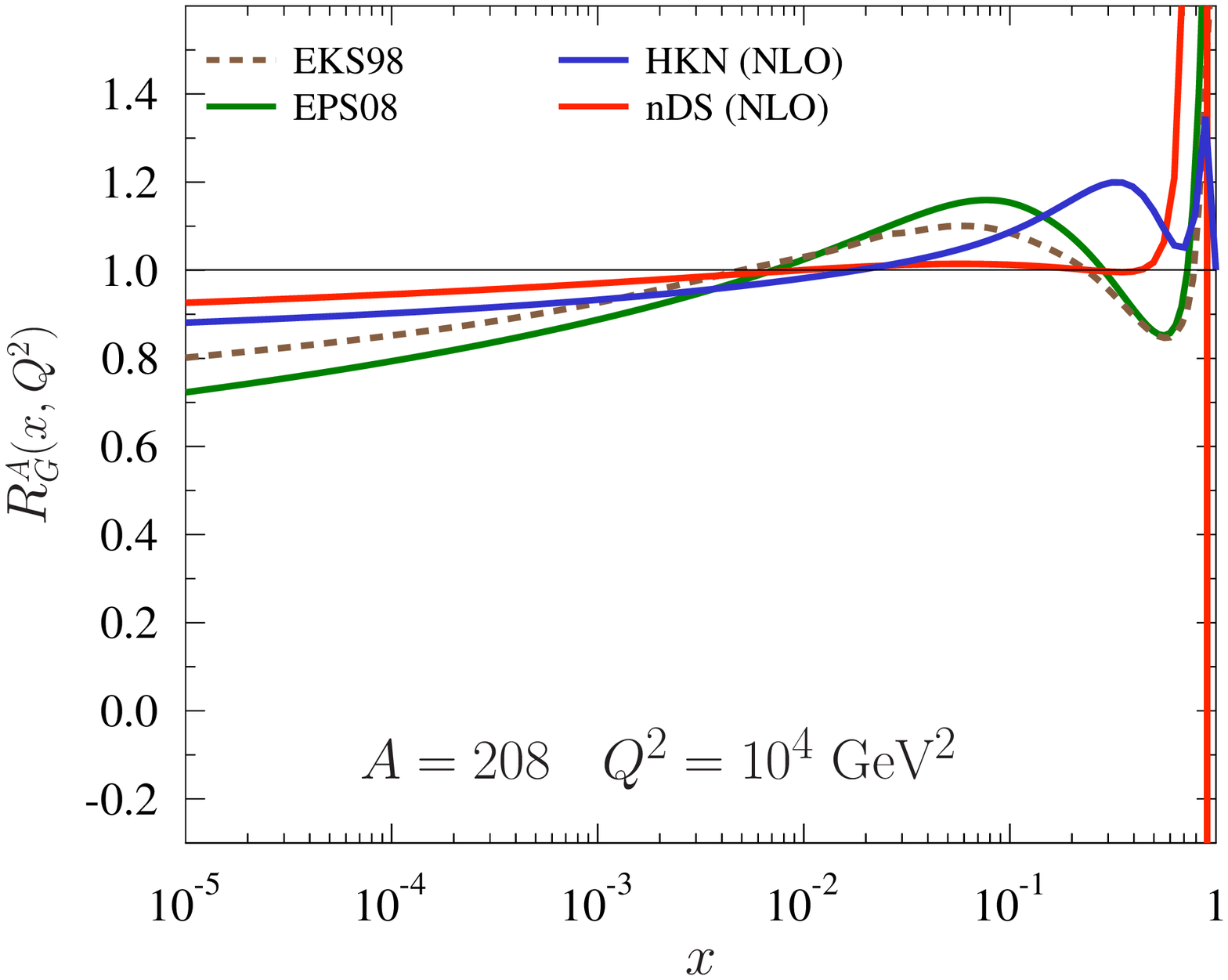}
\end{minipage}
\caption[]{\small Evolution of the gluon ratios $R_g^A(x,Q^2)$ for the parametrizations of the different groups, EKS98, EPS08, HKN, nDS. The differences, present at the smallest virtualities, quickly disappear with increasing $Q^2$.}
\label{fig:gluonsq2}
\end{figure}

The main difficulty in the global fit analyses is to find an initial condition for the evolution (point 1. above) which is flexible to avoid biasing the result as much as possible but with a small enough number of free parameters so that a minimization procedure can be performed. This problem is common to the free proton analyses, but more severe for nuclei due to the much more limited number of experimental data available, specially in the small-$x$ region. In practice, some assumptions are imposed for those kinematical regions and/or flavours less constrained by the data. For example, a common assumption is to take for the valence quarks $R_{u_V}^A(x,Q^2_0)=R_{d_V}^A(x,Q^2_0)\equiv R_V^A(x,Q^2_0)$ or for the sea quarks $R_{\bar u}^A(x,Q^2_0)=R_{\bar d}^A(x,Q^2_0)=R_{\bar s}^A(x,Q^2_0)\equiv R_S^A(x,Q^2_0)$ \footnote{These assumptions are only taken at the initial scale $Q_0^2$, but the equalities do not hold in general for larger virtualities as the evolution of each flavour is different.}. In general, the different analyses give similar results for the ratios $R_i^A(x,Q^2)$ for those regions in which experimental data exist, but differ otherwise. In Fig. \ref{fig:pdfscomp} different sets for the ratios of PDFs of protons inside a lead nucleus over those in a free proton are plotted. The ratios of the different groups are rather similar for valence quarks: well constrained by DIS data except at small-$x$; and sea quarks: constrained by DIS and DY data except for very small and large-$x$. The case of the gluons is, however, less satisfactory as very different parametrizations (specially in the  small-$x$ region) lead to similar descriptions of the available DIS and DY data. In order to improve the situation, additional experimental data able to constrain the gluon distributions need to be included in the analyses. For the first time, dAu data on inclusive high-$p_T$ particle production from RHIC have been included in Ref.  \cite{Eskola:2008ca}. In particular, the strong shadowing of gluons found in this analysis can be traced back to the strong suppression of the forward rapidity negative particle yields measured by BRAHMS. It is worth emphasizing, however, that although the ratios are very different for the lower scales, they become more and more similar with evolution. In particular, the nuclear effects for the gluons at a virtuality of $\sim$ 100 GeV are very similar for all the groups as showed in Fig. \ref{fig:gluonsq2}.

These global fit analyses allows one also to compute the uncertainties for a given observable associated the nPDFs. The procedure, adapted from the free proton analyses, consists in computing the eigenvalues corresponding to the Hessian matrix $H$ defined as 
\begin{equation}
H_{i,j}=\frac{1}{2}\frac{\partial^2 \chi^2}{\partial z_i\partial z_j}{ \Big |_{z_i=z_i^0;\,  z_j=z_j^0}}\, ,
\label{eq:hessian}
\end{equation}
where $\{z_i^0\}$ corresponds to the values of the parameters which minimize $\chi^2$. For $N$ free parameters, the same number of eigenvalues $\{\bar z_i\}$ is obtained, each one defining one eigenvector. These eigenvectors are computed by setting all the parameters  $\{\bar z_i\}$ in the new basis to the values at the minimum $\{\bar z^0_i\}$ except one, $\bar z_k$. In order to compute the uncertainty, two new sets of nPDFs are then computed for $z_k=z_k^0+\delta z_k$ and $z_k=z_k^0-\delta z_k$, where $\delta z_k$ depends on the confidence level required. In this way, $2N$ different sets of nPDFs are defined. To compute the error band associated to a given observable, for example the jet cross-section in heavy-ion collisions at the LHC, one would compute this observable for the $2N$ different sets of nPDFs and the results added in quadrature. A computer code for practical applications including the nPDF error estimates is available in Ref. \cite{Eskola:2009uj}. 

\subsection{Non-linear evolution: saturation of partonic densities}

The saturation of partonic densities described in Section \ref{sec:satur} finds its natural implementation in the presence of non-linear terms in the evolution equations. The first non-linear terms computed were the leading higher-twist corrections to the DGLAP evolution equations \cite{Gribov:tu,MQ}

\begin{equation}
\frac{\partial xg(x,Q^2) }{\partial \log Q^2} 
=   \frac{\partial xg(x,Q^2) }{\partial \log Q^2}\bigg|_{\rm DGLAP}
  - \quad \frac{9\pi}{2} \frac{\alpha_s^2}{Q^2} 
    \int_x^1 \frac{dy}{y} \frac{1}{\pi R^2}[yg(y,Q^2)]^2,
\label{eq:mq}
\end{equation} 
where the first term is the standard DGLAP result \cite{DGLAP}, linear in the PDFs, and $R\simeq1$~fm, the radius of the proton. The approach to the problem is, however, different nowadays and a generalization of the BFKL equation \cite{Kuraev:1977fs,Balitsky:1978ic} is normally used in phenomenological approaches\footnote{See, however, Ref.\cite{Eskola:2002yc} for a recent phenomenological application of Eq. (\ref{eq:mq}).} instead of the DGLAP equation. Several reasons exist for this preference, in particular, the small-$x$ evolution is theoretically enhanced by terms in $\alpha_s\log x$, which are the ones resummed by BFKL. 

\begin{figure}
\begin{center}
\includegraphics[scale=0.5]{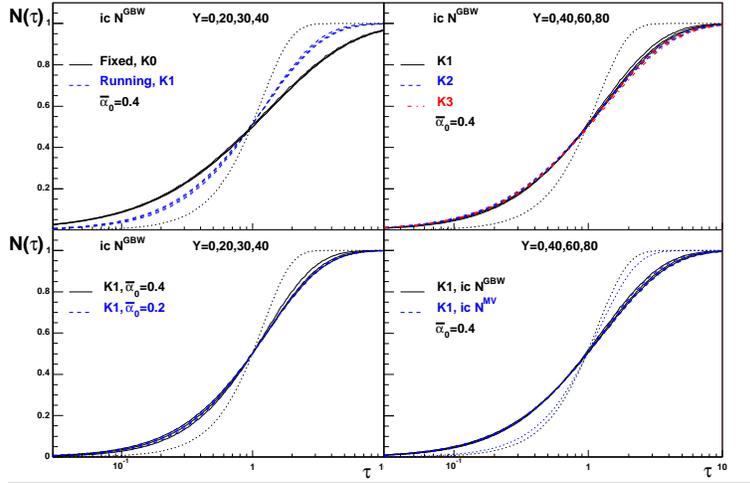}
\end{center}
\caption{Scaling solution of BK equations (\protect\ref{eq:bk}) for different rapidities and initial conditions for the evolution. The scaling variable is defined as $\tau=rQ_{\rm sat}(Y)$. Figure from Ref. \protect\cite{Albacete:2004gw}.}
\label{fig:scalingBK}
\end{figure}

The most widely used evolution equation in phenomenological analyses is the Balitsky--Kovchegov  (BK) \cite{Balitsky:1995ub,Kovchegov:1999yj} equation. It gives the evolution with rapidity 
$Y=\ln{(s/s_0)}=\ln{(x_0/x)}$ of the
scattering
probability $N(\vec{x},\vec{y},Y)$
of a $q\bar q$
dipole with a hadronic target, where $\vec{x}$ ($\vec{y}$) is
the position of the
$q$ ($\bar q$) in transverse space with respect to the centre of the target.
Defining
\begin{equation}
\vec{r}=\vec{x}-\vec{y},\ \ \vec{b}={\vec{x}+\frac{\vec{y}}{2}}\,,\ \ 
\vec{r}_1=\vec{x}-\vec{z},\ \ 
\vec{r}_2=\vec{y}-\vec{z} \, ,
\label{eq:coord}
\end{equation}
and neglecting the impact parameter dependence (which is justified
for $r\ll b$), the BK equation reads ($r\equiv |\vec{r}|$)
\begin{equation}
\frac{\partial N(r,Y)}{\partial Y}=\frac{\alpha_sN_c}{\pi}\int \frac{d^2z}{2\pi}\, \frac{r^2}{r_1^2r_2^2}
\left[N(r_1,Y)+N(r_2,Y)-N(r,Y)-N(r_1,Y)N(r_2,Y)\right]\, .
\label{eq:bk}
\end{equation}
The coupling constant is fixed and the kernel is conformally invariant --- this means, in particular that the solutions present geometric scaling. The NLO corrections to this formula are not known, but a part of them have recently been considered in Refs. \cite{Balitsky:2006wa,Kovchegov:2006vj,Albacete:2007yr}. In (\ref{eq:bk}) the first three terms correspond to the linear, BFKL, equation. They lead to an exponentially increasing scattering probability
\begin{equation}
N_{\rm linear}\simeq x^{-\lambda}\ \ \ \ \mbox{with}\ \ \ \  \lambda=\frac{\alpha_sN_c}{\pi} 4\ln 2\, .
\label{eq:bfkl}
\end{equation}
So, the scattering probability becomes larger than 1 for small enough $x$, violating the unitarity bound which ensures probability conservation. This deficiency of the BFKL equation is rectified by the non-linear contribution which ensures $N(r,Y)\leq 1$ for all values of $r$ and $Y$. The restoration of unitarity is at the heart of the saturation approaches and calls for non-linear terms in the evolution equations. These non-linearities correspond to the gluon fusions in the intuitive picture we provided in Section \ref{sec:satur}. Numerical solutions of the BK equations \cite{Albacete:2004gw} showing the property of geometric scaling are plotted in Fig. \ref{fig:scalingBK}. Similar computations including the running of the coupling \cite{Balitsky:2006wa,Kovchegov:2006vj,Albacete:2007yr} can be found in Ref. \cite{Albacete:2007yr}. A recent fit of HERA data using the BK equation with running coupling made available the corresponding scattering probability for computation of other observables \cite{Albacete:2009fh}.

\subsection{The charmonium suppression}

\begin{figure}
\begin{center}
\begin{minipage}{0.3\textwidth}
\begin{center}
\includegraphics[scale=0.5]{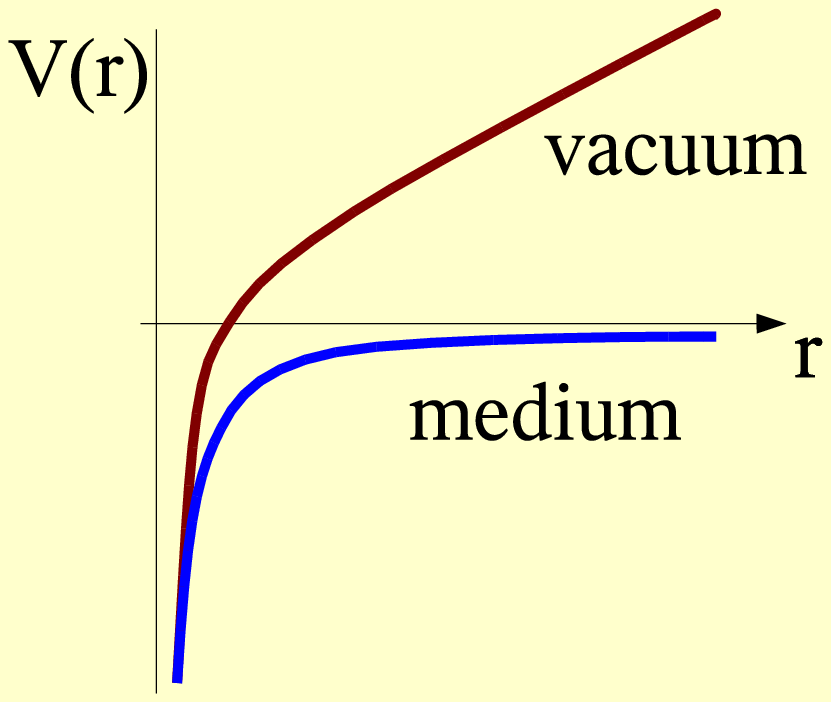}
\end{center}
\end{minipage}
\begin{minipage}{0.3\textwidth}
\begin{center}
\includegraphics[scale=0.2]{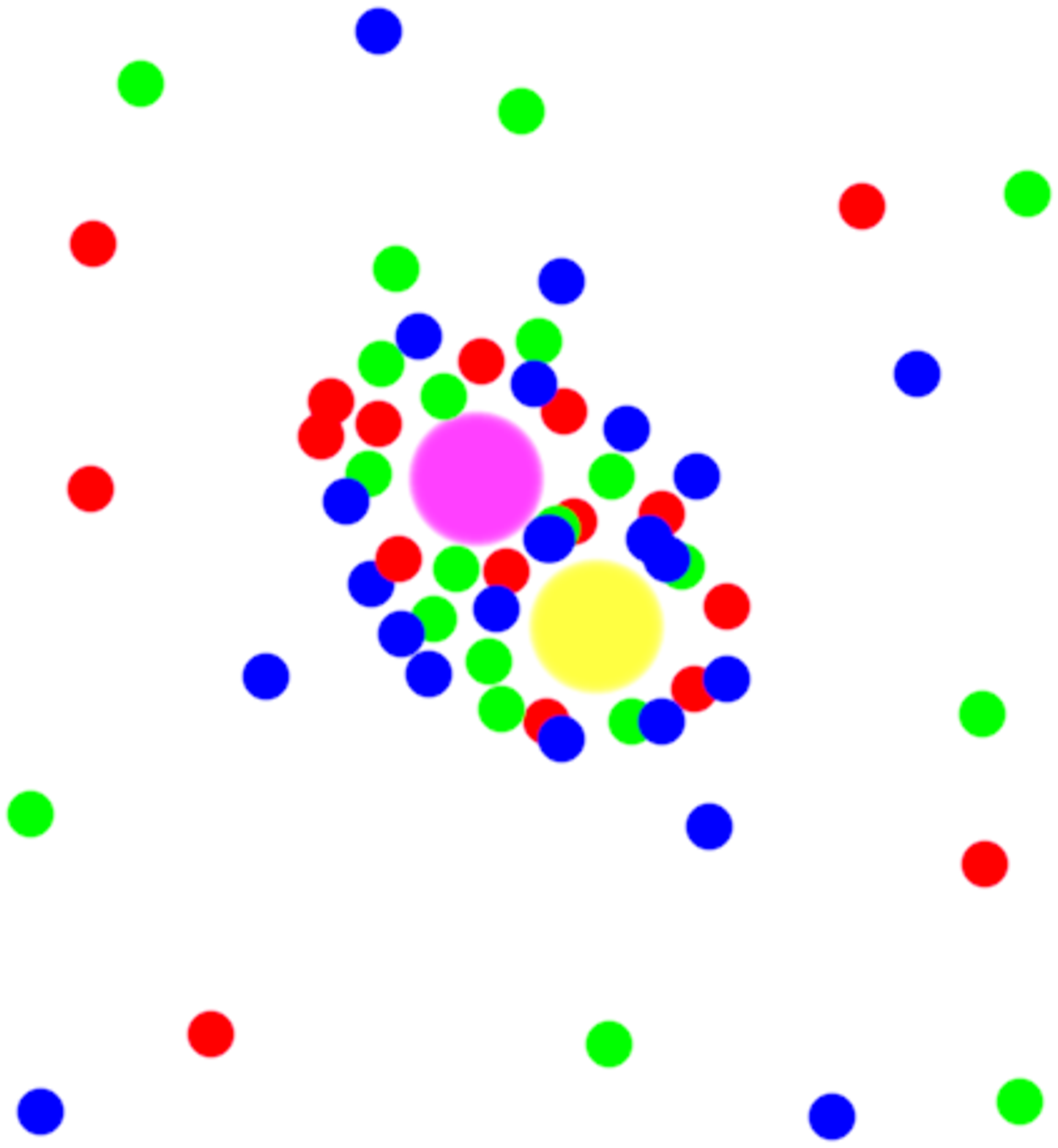}\\
Vacuum
\end{center}
\end{minipage}
\begin{minipage}{0.3\textwidth}
\begin{center}
\includegraphics[scale=0.2]{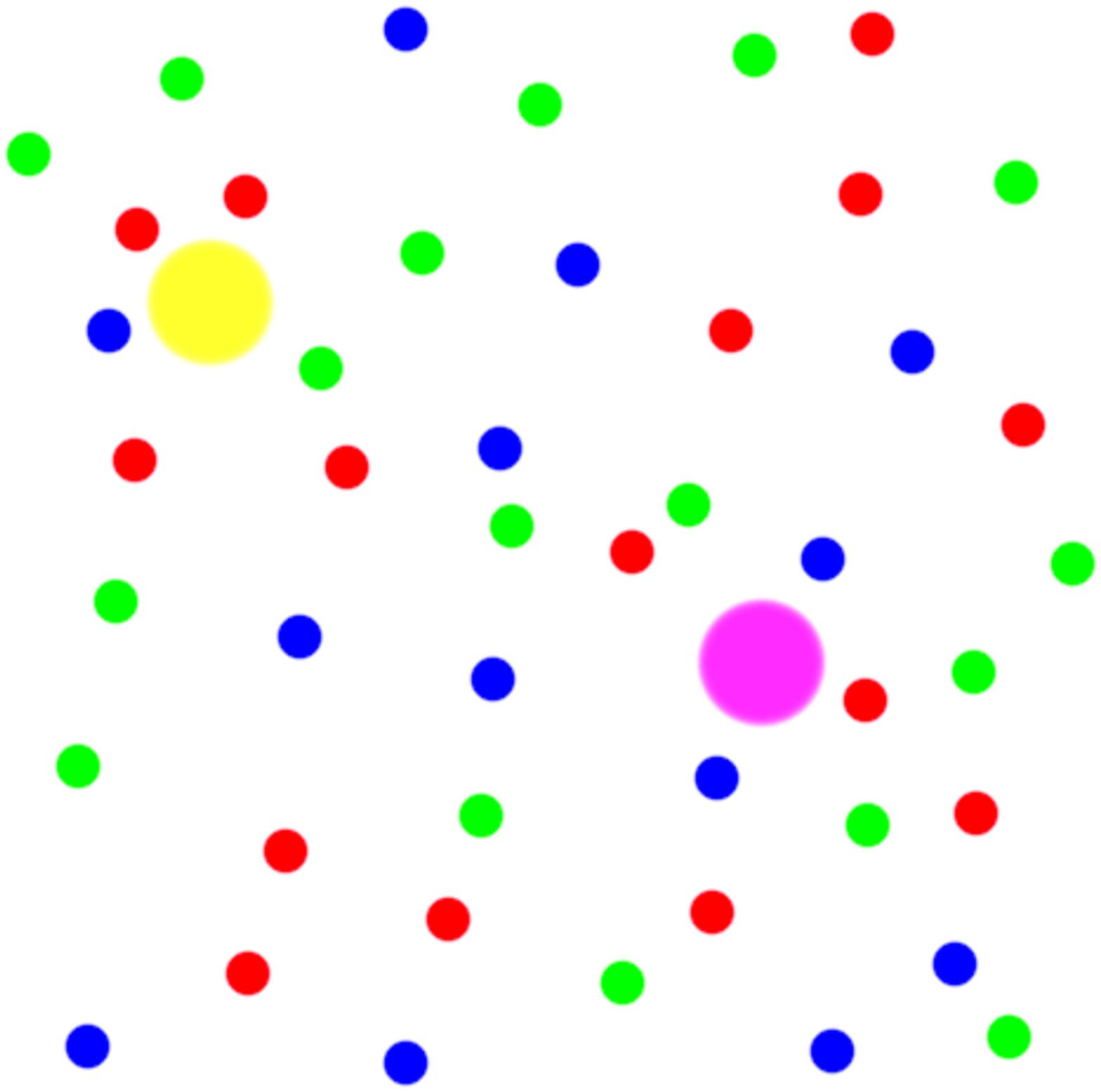}\\
Medium
\end{center}
\end{minipage}
\end{center}
\caption{ (Left) Pictorial representation of the potential in the vacuum, Eq. (\ref{eq:pot}), and the corresponding screened potential in the medium; (Centre) A confined $c\bar c$ pair in the vacuum forming a $J/\Psi$; (Right) In the case where the $c\bar c$ pair is in a thermal medium, the $J/\Psi$ is dissolved.}
\label{fig:jpsisup}
\end{figure}

The first of the hard probes experimentally accessible was the $J/\Psi$ suppression, proposed in 1986 by Matsui and Satz \cite{Matsui:1986dk} and first measured one year later at the CERN SPS in oxygen--uranium collisions \cite{Abreu:1988tp}. The original formulation of the problem considered the modifications of the potential (\ref{eq:pot}) in the case where the heavy quarks are inside a medium. The small-distance part of the potential can be computed perturbatively and the long-range Coulomb $-\alpha_{\rm eff}/r$ is screened to a short-range Yukawa potential, $-\alpha_{\rm eff}\exp\{-r/r_D(T)\}/r$. The modification of the confining linear term in (\ref{eq:pot}) is not that simple\footnote{This is a matter of current debate, see, for example, Ref. \cite{Mocsy:2008eg} and references there.}, but in first approximation, a deconfined medium should produce $\sigma\to 0$. So, a possible modified potential reads (see Fig. \ref{fig:jpsisup})
\begin{equation}
V(r)\simeq -\frac{\alpha_{\rm eff}}{r}\exp\{-r/r_D(T)\}\, .
\label{eq:screen}
\end{equation}
With this set-up, Matsui and Satz found the probability of survival for a $J/\Psi$ after a thermal deconfined state is formed to be very small. The conclusion was that the suppression of the $J/\Psi$ (as well as other charmonium and bottonium states) signals unequivocally the formation of a QGP.

The interpretation of the observed $J/\Psi$ suppression is, however, difficult because the theoretical control over the modification to both hot and cold nuclear matter is not precise enough. Indeed, $J/\Psi$ is already suppressed in hadron--nucleus collisions where the formation of a deconfined state is not possible owing to the small energy densities created. The  charmonium states are fragile enough to be destroyed by interaction with the surrounding nuclear matter once produced inside the nuclei. The proton--nucleus programme at the SPS was essential to fix the benchmark for this {\it normal nuclear absorption}. Although the understanding of the $J/\Psi$ suppression is still not satisfactory, a good description of the pA data is provided by a simple probabilistic approach in which a $J/\Psi$ produced at a given longitudinal position $z_0$ inside the nucleus travels essentially in a straight line with subsequent scatterings at $z_1<z_2<...$ If the absorption or breakup cross-section for one elementary $J/\Psi-$nucleon cross-section is $\sigma^{\rm abs}$ the corresponding survival probability at a given impact parameter can be computed in the Glauber model \cite{Capella:1988ha}.

Let us take the case of pA collisions for simplicity. The $J/\Psi$ is produced at an impact parameter $b$ and longitudinal position $z$ inside the nuclei with a proton--nucleon $J/\Psi$ production cross-section $\sigma_\psi$. Were the $J/\Psi$ not to interact with matter (as is the case for Drell--Yan dileptons, for example) the cross-section would be
\begin{equation}
{\sigma^{pA\to J/\Psi}_0}(b) =\int_{-\infty}^{\infty}dz\,A\rho_A(z,b)\sigma_\psi=AT_A(b)\sigma_\psi
\label{eq:jpsi0}
\end{equation}
and the collisional scaling is recovered as it should be for hard processes. The $J/\Psi$, however, reinteracts with the nuclear matter while travelling through it with an {\it absorption cross-section}, $\sigma_{\rm abs}$. Repeating the arguments of Section \ref{sec:glauber}, the probability that $n$ $J/\Psi+N$ interactions take place after the $J/\Psi$ is produced at position $z$ is
\begin{equation}
P_n^\psi(z,b)=\begin{pmatrix}A\\n\end{pmatrix}\left[1-\int_z^\infty dz\,\rho_A(z,b)\sigma_{\rm abs}\right]^{A-n}\left[\int_z^\infty dz\,\rho_A(z,b)\sigma_{\rm abs}\right]^{n}\, .
\end{equation}
So, the survival probability corresponds to the case $n=0$ and the cross-section (\ref{eq:jpsi0}) is modified to\footnote{We assume here that $\sigma_\psi$ is small to neglect terms ${\cal O}(\sigma_\psi^2)$, i.e. only one $J/\Psi$ is produced in the nucleus}
\begin{equation}
\sigma^{pA\to J/\Psi}(b) =\int_{-\infty}^{\infty}dz\,A\rho_A(z,b)\sigma_\psi\exp\left\{\int_z^\infty dz\,A\rho_A(z,b)\sigma_{\rm abs}\right\}\, ,
\label{eq:jpsi}
\end{equation}
where we have used again the approximation of large-$A$. The integral in $z$ can be done by expanding the exponent and using the relation
\begin{equation}
\int_{-\infty}^\infty dz_1 \rho(z_1,b)\int_{z_1}^\infty dz_2 \rho(z_2,b)...\int_{z_n}^\infty dz_n \rho(z_n,b)=\frac{1}{n!}
\left[\int_{-\infty}^\infty dz \rho(z,b)\right]^n
\label{eq:recurs}
\end{equation}
back and forth to obtain
\begin{equation}
\sigma^{pA\to J/\Psi}(b) =\frac{\sigma_\psi}{\sigma_{\rm abs}}\left[1-\exp\left\{-\sigma_{\rm abs}AT_A(b)\right\}\right]\, .
\label{eq:jpsib}
\end{equation}
The corresponding expression for AB collisions can be obtained using similar methods to obtain, again in the optical approximation,
\begin{equation}
\sigma^{AB\to J/\Psi}(b) =\frac{\sigma_\psi}{\sigma_{\rm abs}^2}\int d{\bf s}\left[1-\exp\left\{-\sigma_{\rm abs}AT_A({\bf s})\right\}\right]\left[1-\exp\left\{-\sigma_{\rm abs}BT_B({\bf b-s})\right\}\right]\, .
\label{eq:jpsiab}
\end{equation}

\begin{figure}
\begin{minipage}{0.5\textwidth}
\begin{center}
\includegraphics[scale=0.4]{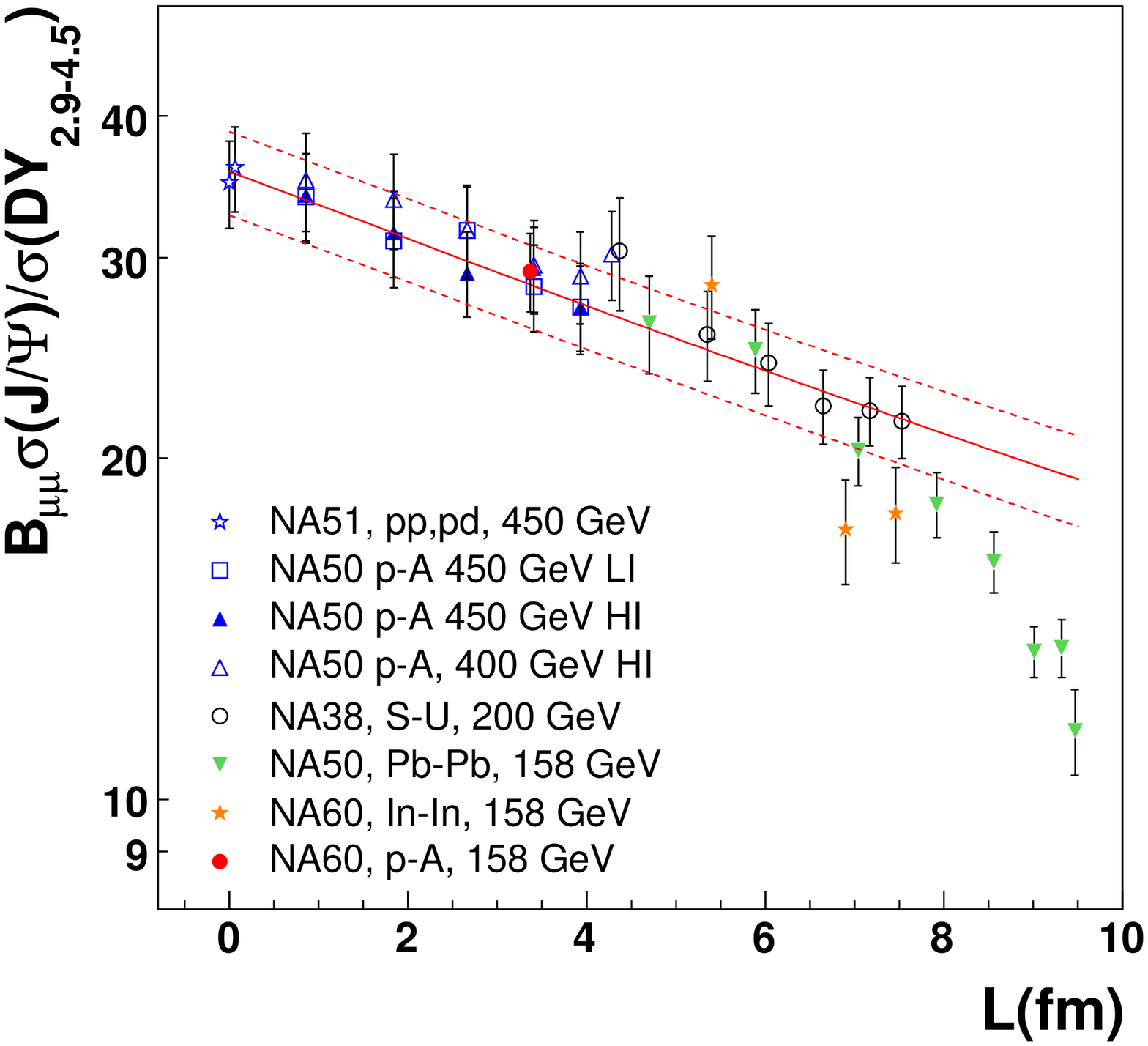}
\end{center}
\end{minipage}
\begin{minipage}{0.5\textwidth}
\begin{center}
\includegraphics[scale=0.4]{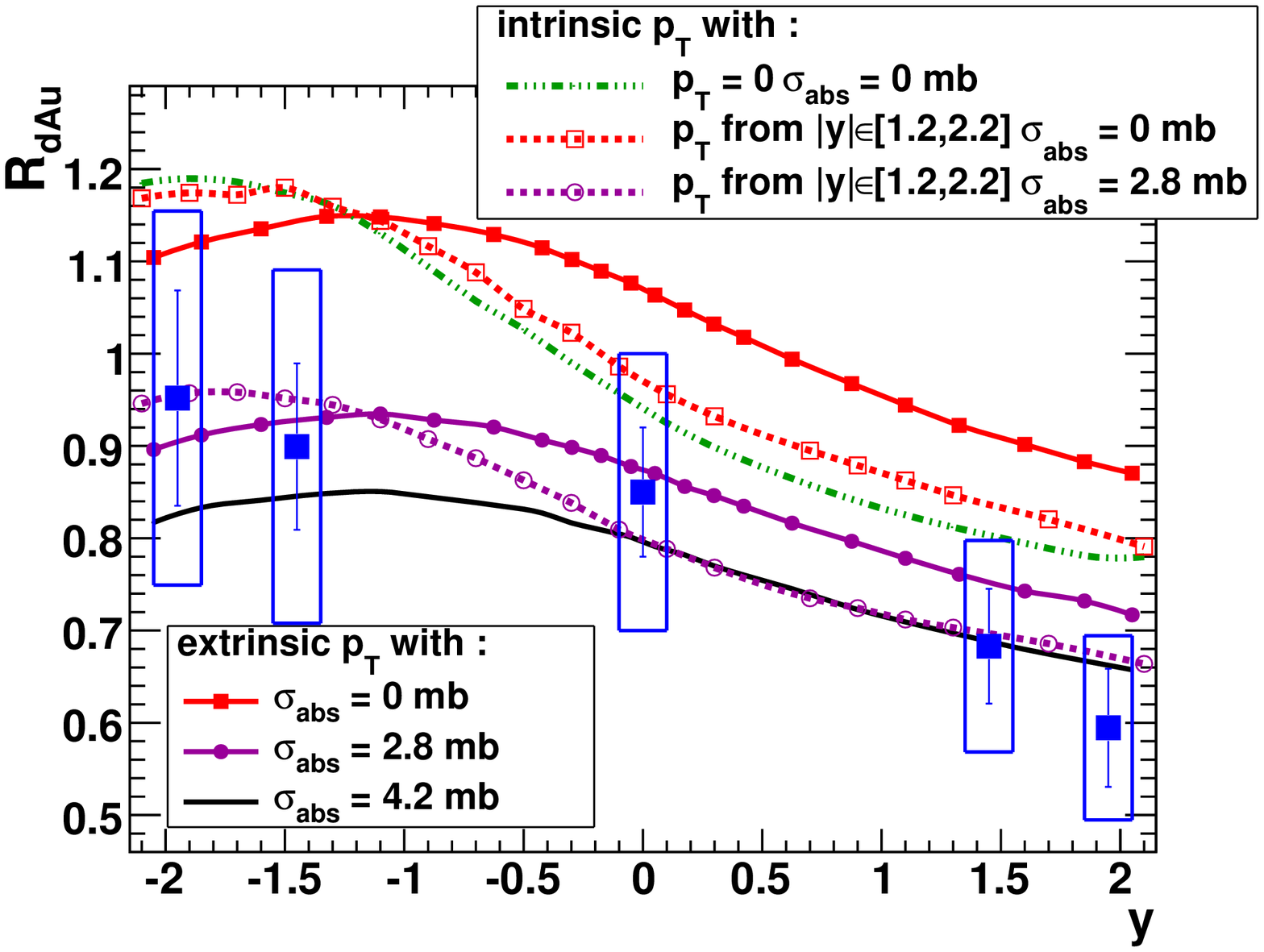}
\end{center}
\end{minipage}
\caption{(Left) $J/\Psi$ suppression in pA and AB collisions for different systems and energies measured at CERN SPS \cite{Abreu:1999qw,Alessandro:2006jt,Arnaldi:2007zz}; (Right) $J/\Psi$ suppression in dAu collisions at RHIC as a function of rapidity compared to different models of $J/\Psi$ production using EKS98 nuclear PDFs, figure from Ref. \protect\cite{Ferreiro:2008wc}.}
\label{fig:jpsi}
\end{figure}

Equations (\ref{eq:jpsib}) and (\ref{eq:jpsiab}) provide a good description of the suppression observed at SPS in pA, SU and peripheral InIn and PbPb collisions. This led to the interpretation of the {\it anomalous $J/\Psi$ suppression} --- that which is not described by cold nuclear matter effects (\ref{eq:jpsiab}) --- as due to the formation of a QGP \cite{Abreu:2000ni}, see Fig. \ref{fig:jpsi}. At RHIC, however, where a larger suppression was expected due to the larger temperatures and densities of the medium, data point to a similar magnitude of the effect \cite{Adare:2006ns}. Other effects like shadowing have also been studied, see Fig. \ref{fig:jpsi}, but the interpretation of the data is still under debate. One of the main difficulties is the lack of a good benchmark, both for proton--proton collisions, where the actual mechanism of $J/\Psi$ production is not known, and proton--nucleus collisions where the relative magnitude of the different effects (shadowing, nuclear absorption, etc.) cannot be fixed by present experimental data\footnote{In fact, even the validity of Eqs. (\ref{eq:jpsib}) and (\ref{eq:jpsiab}) needs to be assessed in view of the simplicity of the probabilistic approach used in their derivation. In this respect, notice that the probabilistic approach can be recovered by the low-energy (incoherent) limit of a simplified field theoretical calculation including formation time effects \cite{Braun:1997qw}.}. New data from RHIC and the LHC is expected to further clarify the situation, including the study of other charmonium ($\chi_c$, $\Psi'$) and bottonium ($\Upsilon$, $\Upsilon'$...) states as well as open charm and bottom, see, for example, Ref.  \cite{deCassagnac:2009dr} and references therein.

\subsection{Jet quenching}

\label{sec:jetquenching}

The original idea to characterize the properties of the medium produced in hadronic collisions by the suppression of particles at large transverse momentum is due to  Bjorken \cite{Bjorken:1982tu}. In the simplest view, a parton produced in a hard process inside the medium loses energy when travelling though the produced matter; by elastic scatterings in Bjorken's original proposal. This energy loss leads to a strong suppression of the high-$p_T$ particle yields due to the steeply-falling spectrum in this region of phase space; a 10\% energy loss translates into a factor of $\sim 2$ suppression of the yields.

The cross-section to produce a hadron $h$ at a given $p_T$ can be computed using Eq. (\ref{eq:hard}) which reads explicitly, see, for example, Ref. \cite{Eskola:2002kv} for details
\begin{equation}
\frac{d\sigma^{AB\to h}}{dp_T^2 dy}=\sum_{i,j,k=q,\bar q,g}\int {dy_2}\frac{dz}{z^2}x_1f_{i/A}(x_1,Q^2)x_2f_{j/B}(x_2,Q^2)\frac{d\hat\sigma^{ij\to kl}}{d\hat t} D_{k\to h}(z,\mu_F^2)\, .
\label{eq:hardexp}
\end{equation}
The fragmentation functions $D_{k\to h}(z,\mu_F^2)$ describe the hadronization of a parton $k$ into a hadron $h$ with fraction of momentum $z$. In the vacuum, these quantities evolve in $\mu_F^2$ with DGLAP evolution equations in a fashion similar to the PDFs. They are also obtained by global fits like the ones described above. These quantities are, however, modified in the case where this fast parton traverses a medium. A complete theoretical understanding of this modification is not yet known and a simple picture including only the energy loss of the leading particle is usually assumed; an excellent recent review on the fragmentation functions in the vacuum and medium is available in Ref. \cite{Arleo:2008dn}:
\begin{equation}
D^{\rm med}(x,t)=\int \frac{d\epsilon}{1-\epsilon}P(\epsilon)D^{\rm vac}\left(\frac{x}{1-\epsilon},t\right)\, .
\label{eq:mffqw}
\end{equation}
Here, $P(\epsilon)$ is the probability that the particle loses a fraction $\epsilon$ of its energy. For the moment we have not specified the nature of the energy loss mechanism, we have just assumed that the effect is additive to the usual DGLAP evolution in a probabilistic way: energy loss does not change the evolution.

Asymptotically, the main source of energy loss of a particle traversing a medium is radiation  \cite{Peigne:2008wu}. In the next section we shall present the calculation of the medium-induced gluon radiation, the generalization of bremsstrahlung for gluons produced in a finite size medium. A main feature of the corresponding spectrum is the presence of a natural collinear cut-off given by the generalization to QCD of the Landau--Pomeranchuk--Migdal effect. The typical singularities, which in the vacuum are resummed in the evolution equations, are then absent, justifying a probabilistic implementation. 

We shall not discuss here the relative importance of different energy loss mechanisms or implementations, but just present some results taking radiative energy loss as the main source of the effect. The usual implementation of the effect assumes an independent gluon emission approximation, which in the eikonal case leads to a Poisson distribution for the energy loss probabilities --- also known as {\it quenching weights} \cite{Baier:2001yt,Salgado:2003gb}\footnote{A code with the numerical values is available at URL \cite{qw-site}.}
\begin{equation}
P(\Delta E)=p_0\,\sum_{k=0}^\infty \frac{1}{k!}\int\left[\prod_{i=1}^k d\omega_i\int_0^{\omega_i}dk_\perp\frac{dI^{\rm med}(\omega_i)}{d\omega dk_\perp}\right]\delta\left(\sum_{j=1}^k \omega_j-\Delta E\right)\hspace{0.2cm}
\label{eq:qwfin}
\end{equation}
where $dI^{\rm med}/d\omega dk_\perp$ is the spectrum of medium-induced gluons,  the normalization is given by the probability of no-energy loss
\begin{equation}
p_0=\exp\left[-\int_0^\infty d\omega\int_0^\omega dk_\perp\frac{dI^{\rm med}}{d\omega dk_\perp}\right] \, ,
\label{eq:p0}
\end{equation}
and $\epsilon=\Delta E/E$. The total distribution contains a discrete part, the term with $k=0$ in (\ref{eq:qwfin}), giving the probability that the particle exits the medium unaffected, and a continuous part for a finite amount of energy loss $\Delta E$
\begin{equation}
P(\Delta E)=p_0\delta(\Delta E)+p(\Delta E)\, .
\label{eq:qwtot}
\end{equation}
All the information about the medium is contained in the medium-induced gluon radiation spectrum $dI^{\rm med}/d\omega dk_\perp$ which will be presented in the next section. In the approximation in which an arbitrary number of scatterings with the medium are allowed, i.e., when the opacity of the medium is large there is a natural parameter, called the transport coefficient, $\hat q$, encoding all the information of the medium except its total length $L$. These two quantities are the only two parameters of the spectrum and the goal of the jet quenching studies is to obtain the value of $\hat q$ which better describes the data once the geometry, $L$, of the medium is known.

The question of the geometry is far from being trivial for the realistic heavy-ion collisions which are made in the laboratory. Although there is a rather good experimental control on the centrality of the collision, the actual energy density profiles of the created medium still present some uncertainties. For the jet quenching phenomenology, the implementation of a geometry as realistic as possible translates into a better determination of the transport coefficient $\hat q$, or the corresponding parameter in other approaches. An example of this is the determination of this parameter performed within the same approach but using different geometries:
\begin{enumerate}

\item Taking a fixed length of $L=6$ fm (roughly the average distance traversed by a parton produced inside a gold nucleus of radius $L\sim 1.2 A^{1/3}$ and travelling perpendicular to the beam direction) the estimated value is $\hat q\simeq 1$ GeV$^2$/fm \cite{Salgado:2003gb}.

\item Taking the same geometry, a cylinder of radius $L=6$ fm,  but now computing the suppression of the yields before averaging in the length, the resulting value is\footnote{The same value is obtained if instead of a cylinder a Woods--Saxon distribution is assumed for the energy density \cite{Dainese:2004te}.} $\hat q\simeq 5\div 15$ GeV$^2$/fm \cite{Eskola:2004cr}.

\item Taking a geometrical profile from a hydrodynamical description of the heavy-ion collision the obtained value is $\hat q\simeq 8$ GeV$^2$/fm  \cite{Armesto:2009zi}. Now, the medium is not static but evolving in time with longitudinal and transverse expansion. In this case, the average value of the transport coefficient depends on the trajectory --- each different path leads to a different average of $\hat q$.

\end{enumerate}

In recent years an attempt was started to fix the medium properties by common fits to several sets of data \cite{Zhang:2007ja, Adare:2008cg,Renk:2006pk,Renk:2006sx,Bass:2008rv,Armesto:2009zi}.

\subsubsection{Hydro meets jet quenching}

The hydrodynamical description of the soft part of the spectrum in heavy-ion collisions makes this approach very appealing as a realistic geometry implementation of the energy density profiles for the jet quenching calculations. The usual approach is to define a local relation between the transport coefficient and one of the properties of the medium which can be accessible by a hydrodynamical model like the energy density  \cite{Baier:2002tc,Renk:2006pk,Renk:2006sx,Bass:2008rv,Armesto:2009zi}
\begin{equation}
\hat q(x,y,\tau)=2\,K\,\epsilon^{3/4}(x,y,\tau).
\label{eq:qhateps}
\end{equation}
Here, for simplicity, it has been assumed that the particle direction is perpendicular to the beam axis. The factor of $2$ in (\ref{eq:qhateps}) is a useful convention, as, taking $K=1$, the estimate of the transport coefficient for an ideal QGP is recovered \cite{Baier:2002tc}\footnote{Notice that although this is only an estimate, all perturbative estimates lead to similar numerical results for similar medium temperatures, $\alpha_s$, etc. so, we take the $K$-factor as a departure from a perturbative estimate of $\hat q$ at leading order.}. 

The results from Refs. \cite{Salgado:2002cd,Salgado:2003gb,Baier:1998yf,Gyulassy:2000gk} allow us to compute for a dynamical medium (in which the transport coefficient decreases as a function of time due to the longitudinal and transverse expansion of the medium) an equivalent static scenario in which the transport coefficient is given by
\begin{equation}
\langle\hat q\rangle\simeq\frac{2}{L^2}\int d\xi\,(\xi-\xi_0)\,\hat q(\xi)\, ,
\label{eq:qhatequiv}
\end{equation}
Using this relation and the local values of the transport coefficient defined by (\ref{eq:qhateps}), the suppression of high-$p_T$ hadrons can be computed using the factorized expression for the cross-section (\ref{eq:hardexp}) and the corresponding medium-modified fragmentation functions (\ref{eq:mffqw}) calculated using the quenching weights (\ref{eq:qwfin}). In this set-up, once $\epsilon(x,y,\tau)$ is known, the only free parameter is then $K$ to be fitted to experimental data. The following results take these energy density profile functions from the hydrodynamical results in Ref. \cite{Hirano:2003pw}; a code with the numerical output is publicly available at URL \cite{hydro-site}.

The first data to be taken into account is the suppression of light partons; we take into account only $\pi^0$ in the analysis to avoid contributions from baryons. The suppression is normally plotted in terms of the {\it nuclear suppression factor}
\begin{equation}
R_{AA}=\frac{d\sigma_{AA}/dp_T^2dy}{N_{\rm coll}d\sigma_{pp}/dp_T^2dy}\, .
\label{eq:raa}
\end{equation}

\begin{figure}
 \begin{center}
 \includegraphics[width=0.7\textwidth]{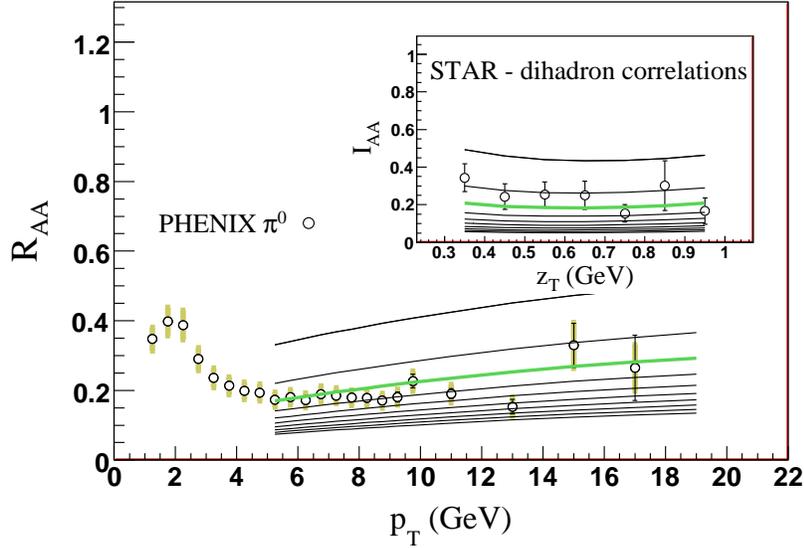}
 \end{center}
\caption{Nuclear modification factors $R_{AA}$ for single-inclusive and $I_{AA}$ for hadron-triggered fragmentation functions for different values of $K=2, 3, 4.. $. The best fit is marked by a wider green line.}
\label{fig:raa1}
\end{figure}

In Fig. \ref{fig:raa1} the results for the suppression of $\pi^0$'s produced at high $p_T$ in central AuAu collisions measured by the PHENIX Collaboration at RHIC \cite{Adare:2008cg} are plotted together with the theoretical calculations for different values of $K$. Highlighted is the value corresponding to $K=4$ which is the one providing the best description for this particular set of data. 

In order to obtain additional constrains to the value of $K$ (or $\hat q$), other observables can be studied. Increasing interest at RHIC is put on the suppression of back-to-back correlations and also on heavy quarks. The latter will be commented on later. The two-particle correlations, on the other hand, are expected to be more sensitive to the medium properties as the surface bias effects would be smaller than in the one-particle inclusive case. In a two-to-two perturbative process like the ones considered here, the two initial partons are produced back-to-back and hadronize independently. This corresponds to a different modification of the fragmentation functions for each of the partons, as they follow different path lengths through the medium. There is, however, no extra free parameter in the calculation. In the insertion of Fig. \ref{fig:raa1} the suppression of the back-to-back azimuthal correlations measured in central AuAu collisions by the STAR  Collaboration at RHIC \cite{Adams:2006yt} are compared with the theoretical curves for the same values of $K$. This is a check that these two sets of experimental data can be described with the same medium parameter $K\simeq 4$. The corresponding definitions are
\begin{equation}
I_{AA}=\frac{D_{AA}(z_T,p_T^{\rm trig})}{D_{pp}(z_T,p_T^{\rm trig})}\, ,
\label{eq:iaa}
\end{equation}
where the {\it hadron-triggered fragmentation function} is defined as \cite{Wang:2003mm}
\begin{equation}
D_{AA}(z_T,p_T^{\rm trig})\equiv p_T^{\rm trig}\frac{d\sigma_{AA}^{h_1h_2}/dy^{\rm trig}dp_T^{\rm trig}dy^{\rm assoc}dp_T^{\rm assoc}}{d\sigma^{h_1}_{AA}/dy^{\rm trig}dp_T^{\rm trig}}\, .
\label{eq:daa}
\end{equation}
Here, $z_T=p_T^{\rm assoc}/p_T^{\rm trigg}$. It should be noticed that the objects defined in this manner are, in general, very different from the fragmentation functions measured, for example, in $e^+e^-$ annihilations at similar virtualities owing to the strong bias produced by triggering in a steeply falling perturbative slope.

The average value of the transport coefficient obtained is, then, $\langle\hat q\rangle\simeq\,K\,2\langle\epsilon\rangle\simeq 8$ GeV$^2$/fm.

\subsubsection{The heavy quarks}

Gluon radiation is suppressed by mass terms in the vacuum by a mechanism called the {\it dead-cone effect}. These mass terms appear in the propagators of the quarks before the radiation point and the corresponding suppression factors are given by the ratio of the massive over massless propagators  \cite{Dokshitzer:1991fd}
\begin{equation}
S_{\rm dead\  cone}^{\rm vac}=\frac{\left(k_\perp^2\right)^2}{(k_\perp^2+x^2M^2)^2}\simeq\frac{1}{\left(1+\frac{\theta_{\rm dc}^2}{\theta^2}\right)^2}.
\label{eq:deadconevac}
\end{equation}
Here, the dead-cone angle $\theta_{\rm dc}^2\simeq M^2/E^2$ defines the angle below which radiation is suppressed. This suppression can be understood intuitively by the inability of the heavy quark to build coherently the wave function of the gluon due to the different velocities of massless and massive particles; in the limiting case $\theta\to 0$ the radiation should cancel as no constructive interference is possible.

In the case of the medium \cite{Dokshitzer:2001zm,Zhang:2003wk,Djordjevic:2003zk,Armesto:2003jh}, the radiation is also suppressed by terms similar to (\ref{eq:deadconevac}) in the propagators, but, interestingly, the LPM suppression is now less effective as the gluon formation time is smaller for a massive quark than for a massless quark: the suppression of the radiation at small angle is smaller, so the radiation is larger. The net effect is a competition between the suppression due to mass terms in the quark propagators and the smaller suppression of collinear gluons due to LPM interference. However, although in some limited regions of phase space the dead cone {\it is filled} \cite{Armesto:2003jh}, the net effect is a suppression of the radiation by mass terms. So, the prediction from the formalism is a smaller energy loss for heavy quarks than for light quarks. This different effect is controlled by mass terms in the medium-induced gluon radiation expressions and implies no new free parameter in the calculations, providing additional constrains to the determination of the medium properties and the nature of the energy loss.

\begin{figure}[t]
\begin{center}
\includegraphics[width=0.7\textwidth]{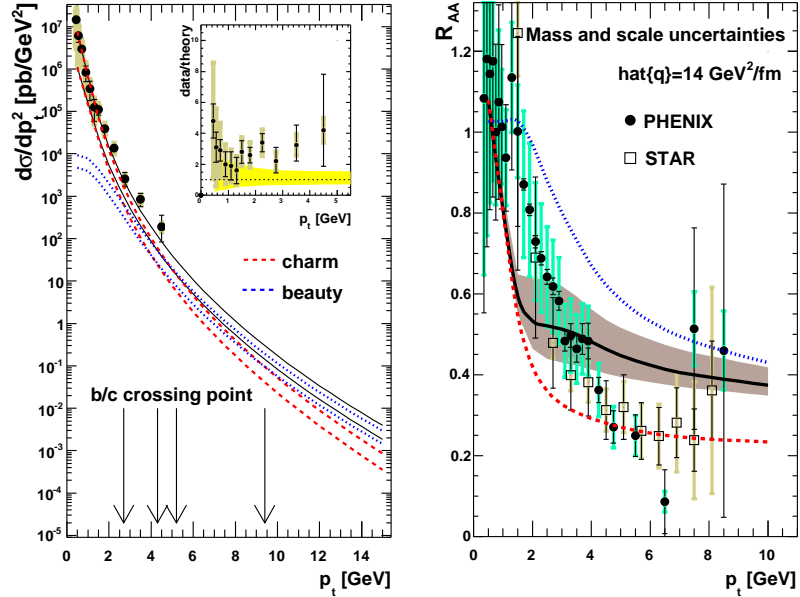}
\end{center}
\caption{(Colour online) (a) Left: Comparison of the FONLL calculation of 
single inclusive electrons~\cite{Cacciari:2005rk} to data from pp collisions at 
$\sqrt{s} = 200$~GeV~\cite{Adler:2005fy}. 
Upper and lower lines are estimates of theoretical uncertainties, obtained by varying 
scales and masses, for details see the text.
(b) Right: The nuclear modification factor $R_{AA}^e$ of electrons in 
central Au--Au collisions for an opacity of the produced QCD matter characterized by
the time-averaged BDMPS transport coefficient $\hat{q} = 14\, {\rm GeV}^2/{\rm fm}$.
The shaded band indicates the theoretical uncertainty of the
perturbative baseline only. Red dashed and blue dotted curves show $R_{AA}^e$
for $c$-quark and $b$-quark decay contributions, respectively. Data taken from Refs.~
\cite{Bielcik:2005wu,Adler:2005xv}. Figure from  Ref. \cite{Armesto:2005mz}.}
\label{fig:hqmatteo}
\end{figure}

Recent attempts to measure the jet quenching of heavy quarks have been made at RHIC \cite{Adler:2005fy,Bielcik:2005wu,Adler:2005xv}. Without a proper vertex reconstruction, the heavy quarks are measured by the semileptonic decays of heavy hadrons into electrons after subtraction of those coming from photons. These {\it non-photonic electrons} originate from the decay of both charm and beauty hadrons and the experimental access to the relative contributions is rather indirect. The theoretical situation is no better and a large indetermination of this quantity is present. State-of-the-art calculations of heavy-quark cross-sections include resummations of large logarithms originated by the mass of the heavy quark in a scheme called Fixed-Order-Next-to-Leading-Log (FONLL) approximation \cite{Cacciari:1998it,Cacciari:2001td}. In Fig. \ref{fig:hqmatteo} the comparison of the non-photonic electrons coming from heavy hadron decays into electrons \cite{Cacciari:2005rk} with experimental data in proton--proton collisions at $\sqrt{s}=200$ GeV at RHIC is shown. The agreement is reasonable within the theoretical uncertainty band computed by varying quark masses and renormalization scales, although the experimental data have the tendency to be underestimated. The relative contributions from $c$ and $b$ quarks are also plotted for the extremes of the bands. When computing the suppression due to medium effects   \cite{Armesto:2005mz}, these uncertainties in the proton--proton benchmark translate into a corresponding uncertainty band for $R_{AA}$ as mass effects are more important for beauty than for charm. This calculations have been performed with a Woods--Saxon profile for the medium, i.e., a static case, but the main conclusions hold equally for the case of a hydrodynamical medium profile.

%

This exercise shows the need for a more precise identification of the charm and the beauty mesons for a correct interpretation of the dynamics underlying energy loss of heavy quarks. It is worth mentioning that the mass effects in the electrons coming from decays of charm hadrons alone are rather small and the suppression is almost mass-insensitive for $p_T^{e}\gtrsim 5$ GeV. The effect is much more important for beauty. So, although taken at face value, the suppression in the experimental data looks to be underestimated by this analysis, a larger contribution of the charm quark would help in reducing this discrepancy   \cite{Armesto:2005mz,Armesto:2005iq}. 

It is also worth mentioning that the formalism explained here is likely to be pushed too far, specially for beauty, and other effects could appear. One of the extra effects which have been advocated in recent years is a large contribution of the {\it collisional energy loss} which would not be negligible for RHIC kinematics \cite{Wicks:2005gt,Peigne:2008nd,Gossiaux:2009hr,Qin:2007rn}. Other possibilities are a larger contribution of the heavy baryons (which have a smaller branching ratio to electrons, explaining part of the suppression) in nuclear collisions at intermediate transverse momentum \cite{Sorensen:2005sm,MartinezGarcia:2007hf}; or non-perturbative modifications of the hadronization of heavy quarks, which owing to the smaller hadronization times and Lorentz factors would {\it hadronize} inside the medium \cite{Adil:2006ra}.

With present experimental data, none of the above descriptions is completely satisfactory and a more precise determination of the mass effect in the heavy-quark suppression would be needed. This identification will be possible with the detectors of the different LHC experiments and also with the upgrades planned for the RHIC experiments.


\subsection{New developments in jet studies in heavy-ion collisions}

A jet is a bunch of particles flying to the detectors in a given limited region of the whole phase space available. This experimental definition corresponds to the production of high-$p_T$ quarks or gluons in hard interactions or in the decay of a very massive particle. In general we can associate large virtualities to these produced partons which then radiate (mainly gluons) to become on-shell and eventually hadronize. So, jets are extended objects corresponding to the large difference in the scales involved: the large $p_T$, defining the typical virtuality of the object, and the hadronization scale ${\cal O}(\Lambda_{\rm QCD})$. In other words, the transverse momentum of the produced parton determines the maximum phase space available for radiation. 

The characterization of jet properties in the vacuum is one of the best tests of QCD. The most favourable experimental condition correspond to $e^+e^-$ annihilation into hadrons. Jets in lepton--proton or proton--proton collisions need a better control over the background. This is specially relevant for the LHC where the background for the {\it underlying event} and/or multiple collisions per crossing in proton--proton collisions will be important. The background will be even larger for central PbPb collisions; peripheral or minimum-bias PbPb collisions will be similar to proton--proton collisions in this respect due to the reduced luminosity for the formed and the slightly smaller centre-of-mass energy.

\subsubsection{The vacuum}

The description of jets in the vacuum is a very well developed subject with a large amount of literature, see, for example, Ref. \cite{ellis}. Here we only provide some heuristic arguments based on the LO description of the effect which is specially relevant for the implementation in Monte Carlo event generators such as PYTHIA \cite{Sjostrand:2006za}, HERWIG \cite{Corcella:2000bw}, or SHERPA \cite{Gleisberg:2008ta}.

The spectrum of gluons radiated with a given energy $\omega$ and transverse momentum $k_T$ by a highly energetic quark or gluon of energy $E$ produced in a hard collision is
\begin{equation}
\frac{dI^{\rm vac}}{dzdk_T^2}=\frac{\alpha_s}{2\pi}\frac{1}{k_T^2}P(z) \rightarrow_{\small z\to 0} \frac{\alpha_s}{2\pi}\frac{1}{k_T^2}\frac{1}{z}\, ,
\label{eq:vacrad}
\end{equation}
where $k_T$ is the transverse momentum of the gluon with respect to the original parton direction and $z=\omega/E$ the fraction of momentum carried by it after the radiation. Equation (\ref{eq:vacrad}) contains infrared ($z\to 0$) and collinear ($\theta\simeq k_T/\omega\to 0$) divergencies. The available phase-space to integrate the radiation is determined by the virtuality $Q^2\simeq p_T^2$ as mentioned above, introducing a cut-off, $\mu^2$ for the collinear divergency: 
\begin{equation}
\int^{Q^2}_{\mu^2}dk_T^2\frac{dI^{\rm vac}}{dzdk_T^2}\sim \alpha_s\log(Q^2/\mu^2)\, .
\label{eq:log}
\end{equation}
So, even when $\alpha_s$ is small, the presence of large logarithms from extended integration regions in phase space makes these higher order terms important and a resummation of an arbitrary number of radiated gluons is necessary. This leads to a gluon emission which is ordered in virtuality. 

An intuitive way of understanding the formulation of the problem is through the {\it radiative decay} analogy \cite{Sjostrand:2006za} in which: i) there is an ordering variable $t$ --- for a radiative decay it is the time, in our case it is the virtuality; ii) the probability that {\it something will happen} at ``time'' $t$ (a decay or a parton branching) is given by a function $f(t)$; iii) there is an additional requirement that something can only happen at time $t$ if it did not happen at earlier times. In these conditions, the probability that nothing has happened, ${\cal N}(t)$,  is given by 
\begin{eqnarray}
\frac{d{\cal N}}{dt}=-f(t){\cal N}(t)\hspace{0.5cm}&\Longrightarrow\hspace{0.5cm}&{\cal N}(t)=\exp\left\{-\int_{t_0}^t f(t')dt'\right\}\, .
\end{eqnarray}
So that, the probability that there is a branching or a decay at time $t$ is just
\begin{equation}
{\cal P}(t)=f(t)\exp\left\{-\int_{t_0}^t f(t')dt'\right\}\, .
\end{equation}
In our case of a parton branching, $f(t)$ is known and the resulting probability to radiate one gluon at a given virtuality $Q^2$ from a parton produced with virtuality $Q^2_{\rm max}$ would be (see also Ref. \cite{ellis} for more details)
\begin{eqnarray}
\frac{d{\cal P}(z,Q^2)}{dzdQ^2}=\frac{dI^{\rm vac}}{dzdQ^2}
\exp\left\{-\int_{Q^2}^{Q^2_{\rm max}}dt\int dz\frac{dI^{\rm vac}}{dzdt}\right\}\equiv \frac{dI^{\rm vac}}{dzdQ^2}\label{eq:1emis0}\Delta\left(Q^2_{\rm max}\right)\Delta\left(Q^2\right)\, .
\label{eq:multvac}
\end{eqnarray}
Notice that removing the virtuality ordering in (\ref{eq:multvac}) would lead to a Poisson distribution for the multiple gluon radiation of the form (\ref{eq:qwfin}).  Writing the spectrum explicitly,
\begin{equation}
d{\cal P}(z,Q^2)=\frac{\alpha_s}{2\pi}\frac{1}{Q^2}P(z)\Delta\left(Q^2_{\rm max},Q^2\right)\,dz\,dQ^2\, .
\label{eq:1emis}
\end{equation}
The term $\Delta(t)$, defined as 
\begin{equation}
\Delta(t)\equiv\exp\left[-\int_{t_0}^t\frac{dt}{t}\int dz\frac{\alpha_s}{2\pi}P(z)\right],
\label{eq:sudakovdef}
\end{equation}
is called the Sudakov form factor and gives the probability that no radiation took place between the two scales $t_0$ and $t$.
This approach can  also be applied to the initial-state radiation for the PDFs. Now the virtuality ordering goes in the opposite direction, increasing from an initial $Q_0^2$ to a given $Q^2$ value. So, if the initial condition for a given PDF is $f(x,Q^2_0)$, its value at $Q^2$ is given by the iteration of (\ref{eq:1emis})
\begin{equation}
f(x,Q^2)=\Delta(Q^2)f(x,Q^2_0)+\int \frac{dt}{t}\frac{\Delta(Q^2)}{\Delta(t)}\int \frac{dz}{z} P(z)f\left(\frac{x}{z},t\right)\, .
\label{eq:evolsud}
\end{equation}
The first term is the contribution in which no resolving radiation corrects the initial value of the PDF between the two scales and the second one gives the evolution due to radiation. Equation (\ref{eq:evolsud}) is the integral representation of the DGLAP equations for only one flavour, which in its more familiar differential form stands as\footnote{The splitting functions in the integral form are, however, regularized through the $+$ prescription whose origin is the virtual corrections when computing the one-gluon emission, see, for example, Ref. \cite{ellis}.}
\begin{equation}
\frac{\partial f(x,Q^2)}{\partial\log Q^2}=\frac{\alpha_s}{2\pi}\int\frac{dz}{z}P(z)f\left(\frac{x}{z},t\right)\, .
\label{eq:dglap}
\end{equation}
So, the DGLAP evolution equations take care of the logarithms (\ref{eq:log}) and the same evolution applies for PDFs and jets in the appropriate kinematical regimes.

The probabilistic approach given by Eq. (\ref{eq:evolsud}) is specially interesting for Monte Carlo implementation. The parton shower algorithms in all of these codes follow the ideas sketched above with some modifications for each particular implementation. We can, in fact, understand these processes as an almost independent gluon emission approximation, in which the presence of an ordering variable breaks the complete independence. The ordering variable we have been presenting above is the virtuality, but other more or less sophisticated ordering variables are also used. For example, HERWIG defines a variable which leads to an ordering in the emission angle of the subsequent partons. This angular ordering encodes colour coherence effects which are absent otherwise. 

The Monte Carlo implementation of the parton shower in event generators follows an iterative procedure; for definiteness, here we take the case of a time-like (final-state) evolution of a produced jet. The basic problem consists in generating the virtuality $t_2$ and fraction of momentum $x_2$ for a branching which takes place after a branching with virtuality $t_1$ and fraction of momentum $x_1$.
\begin{enumerate}
\item First, the virtuality $t_2$ is decided by choosing a random number $\cal R$ and solving 
\begin{equation}
\frac{\Delta(t_2)}{\Delta(t_1)}={\cal R}.
\label{eq:dicet}
\end{equation}
This equation makes use of the probabilistic interpretation of the Sudakov form factor explained above.

\item The next step is to compute the value of the fraction of momentum of the radiated gluon ($x_2$) by solving
\begin{equation}
\int^{x_2/x_1}_\epsilon dz \frac{\alpha_s}{2\pi}P(z)={\cal R'}\int_\epsilon^{1-\epsilon}dz\frac{\alpha_s}{2\pi}P(z)\, ,
\end{equation}
with ${\cal R'}$ another random number and $\epsilon$ an infrared cut-off. 

\item Once the values of $t_2$ and $x_2$ are obtained, checks are made; these could depend on the actual MC implementation. If the branching is possible, a new branching is tried by repeating points 1, and 2, above with the appropriate kinematical variables. The branching stops when some cut-off scale $t_0={\cal O}(1\,\mbox{GeV})$ is reached.

\item The produced partons (quarks and gluons) are colour-connected in a definite manner so that colourless objects can be found. These colourless objects are hadronized by some non-perturbative model into final particles. The hadronization model depends on the actual MC implementation, for example PYTHIA assumes that strings are formed and decay by subsequent breaking, while HERWIG forms colourless clusters of quarks and antiquarks which then decay into the final particles.    

\end{enumerate}

\begin{figure}[t]
\begin{center}
\includegraphics[angle=90,width=0.6\textwidth]{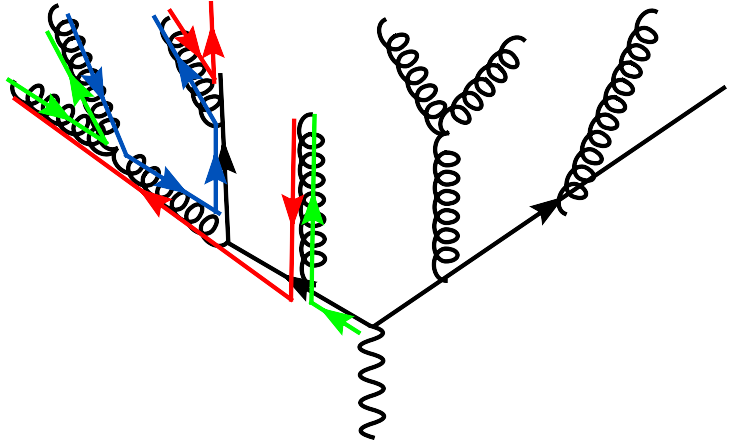}
\end{center}
\caption{Result of the parton shower algorithm explained in the text for the hadronization of a quark--antiquark pair produced in the decay of a virtual photon. The colour flow in only one of the hemispheres is depicted as colour lines.}
\label{fig:jet}
\end{figure}

The typical structure of this shower iterative procedure can be seen in Fig. \ref{fig:jet} where a virtual photon produces a $q\bar q$ pair in opposite directions, each of them forming a parton shower. The colour connection between the different partons is depicted as colour lines for only one of the hemispheres.

%

The description of experimental data provided by the main event generators is excellent in those quantities sensitive to the parton shower structure. For this reason, these codes are essential tools in the phenomenological interpretation of the data and constitute, in fact, one of the main parts of the experimental software for data analysis.

\subsubsection{The implementation of the medium effects}

The vacuum radiation process is a very mature subject in which a large amount of work both from theory and experiment has led to a very good knowledge of the effects involved. The counterpart for the medium radiation is, in comparison, in its infancy. Indeed, on both the theoretical and the experimental side, this is a much more recent subject where a lot of progress is expected in coming years. A learning period with dialog between theory and experiment will be needed. A breakthrough in this respect is the first measurement of reconstructed jets in heavy-ion collisions by the STAR Collaboration \cite{jetsHP} which, although suffering from severe experimental uncertainties, paves the road to future progress in the field. 

Several approaches to the medium implementation of parton showers exist at present \cite{Lokhtin:2005px,Zapp:2008gi,Renk:2008pp,Armesto:2008qh,Armesto:2009fj}, none of them being completely satisfactory but addressing different issues in the problem.

There are several effects which could modify the evolution of a parton shower in a medium after the hard process, in particular:

\begin{enumerate}

\item Medium-induced gluon radiation modifying the vacuum radiation structure. As we have seen, this radiation is free of IR or UV divergencies, but is enhanced by powers of the traversed medium length $\alpha_s\,L$. The angular structure is also modified.

\item Non-eikonal corrections to the parton propagation in medium (also known as {\it collisional energy loss})

\item Energy flow from and to the medium --- which could eventually modify the hydrodynamical evolution of the medium.

\item Modification of the non-perturbative hadronization. If the transverse momentum of the particles is large enough, the non-perturbative (slow) processes would take place in the vacuum, so, as a first approximation all models assume vacuum hadronization.

\item Modification of the colour structure of the shower evolution. I.e., the colour structure plotted in Fig. \ref{fig:jet} corresponds to the vacuum, where colour exchanges are only possible within the created particles. In the case where this structure is formed in the medium, colour exchanges with the particles in the medium are also possible, leading to a modified colour structure of the parton shower. 

\end{enumerate}

One of the main issues to be solved is to find the appropriate ordering variable in the presence of a medium. For the vacuum, the space--time picture of the parton shower is, in fact, irrelevant as no external scale exists. The spatial extension of the medium indicates, however, that {\it time} should play a role as an ordering variable; there are, in fact, known interferences between the length of the medium and the radiation, leading to LPM suppression, which are included in the computations of the medium-induced gluon radiation.

A simple implementation of medium effects has recently been proposed which attempts to include in a consistent manner the vacuum and medium contributions to the parton shower \cite{Armesto:2008qh,Armesto:2009fj,Armesto:2007dt,Polosa:2006hb}. In this approach, a medium term is included in the splitting function
 \begin{equation}
P_{\rm tot} (z)= P_{\rm vac} (z)\to
 P_{\rm tot} (z)=P_{\rm vac}(z)+\Delta P(z,t,\hat{q},L,E)\, ,
 \label{eq:modsplit}
 \end{equation}
as given by the medium-induced gluon radiation spectrum by matching the vacuum term, recalling the definition of the medium-induced gluon radiation as the total radiation in the presence of a medium with the vacuum contribution subtracted:
\begin{eqnarray}
  \omega\frac{dI}{d\omega\, d{\bf k}_\perp}
  = \omega\frac{dI^{\rm vac}}{d\omega\, d{\bf k}_\perp}
    + \omega\frac{dI^{\rm med}}{d\omega\, d{\bf k}_\perp}\, .
  \label{2.4}
\end{eqnarray}
The vacuum spectrum defines the (vacuum) splitting function, which at small fraction $x=1-z$ is
%
\begin{equation}
\frac{dI^{\rm vac}}{dz\, d{\bf k}_\perp^2}=\frac{\alpha_s}{2 \pi}
\frac{1}{{\rm k}_\perp^2} P^{\rm vac}(z),\ \ P^{\rm vac}(z)
\simeq \frac{2 C_R}{1-z}\,.
\label{vacsplit}
\end{equation}
This indicates the following matching for the definition of the medium-modification term of the splitting function\cite{Armesto:2007dt,Polosa:2006hb}
\begin{equation}
\Delta P(z,t)\simeq \frac{2 \pi  t}{\alpha_s}\, 
\frac{dI^{\rm med}}{dzdt} ,
\label{medsplit}
\end{equation}
where $\omega=(1-z) E$ and ${\bf k}_\perp^2=z(1-z)t$ are taken in the medium-induced gluon radiation spectrum; parton masses have been neglected.

\begin{figure}[t]
\begin{center}
\begin{minipage}{0.5\textwidth}
\begin{center}
\includegraphics[width=\textwidth]{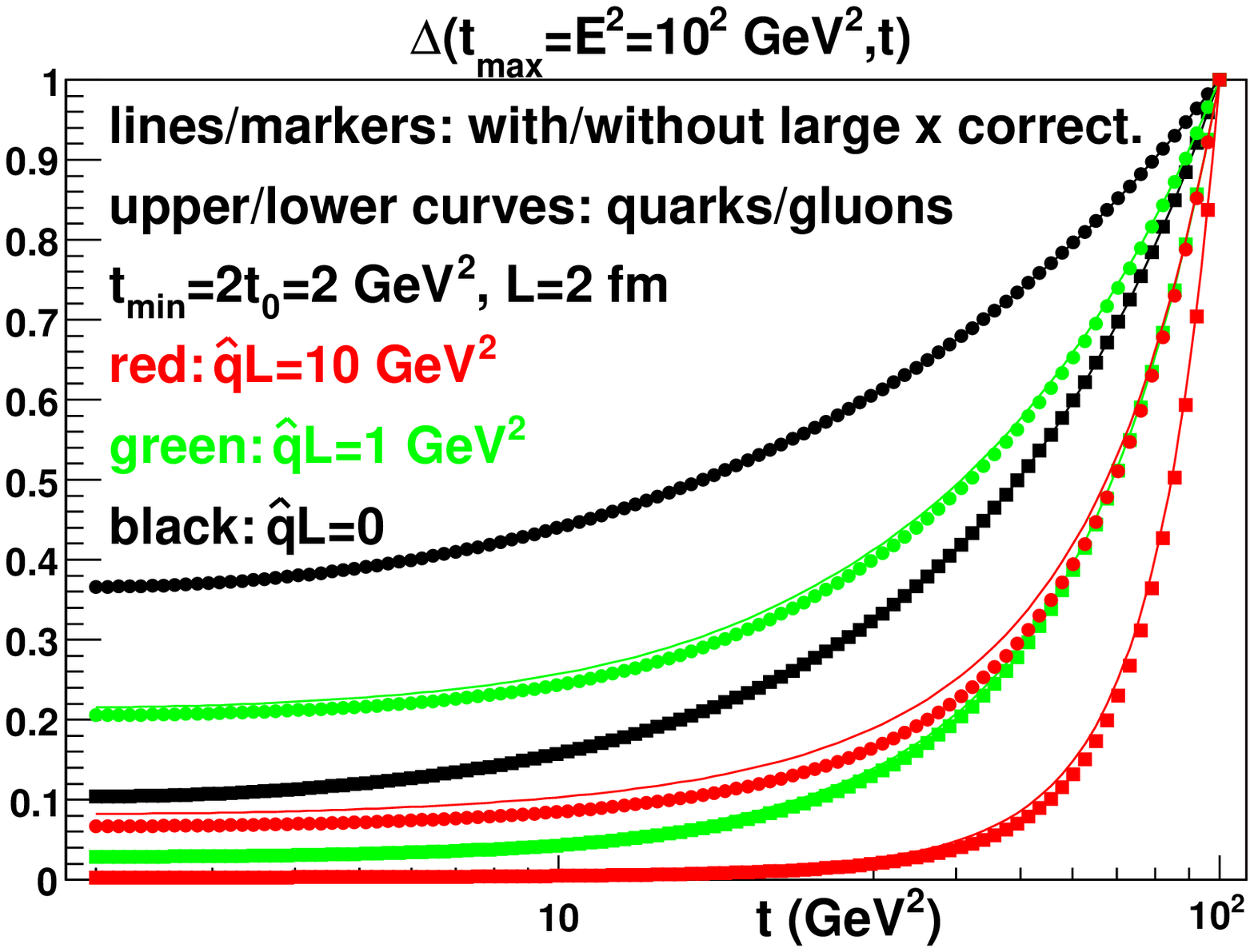}
\end{center}
\end{minipage}
\begin{minipage}{0.5\textwidth}
\begin{center}
\includegraphics[width=\textwidth]{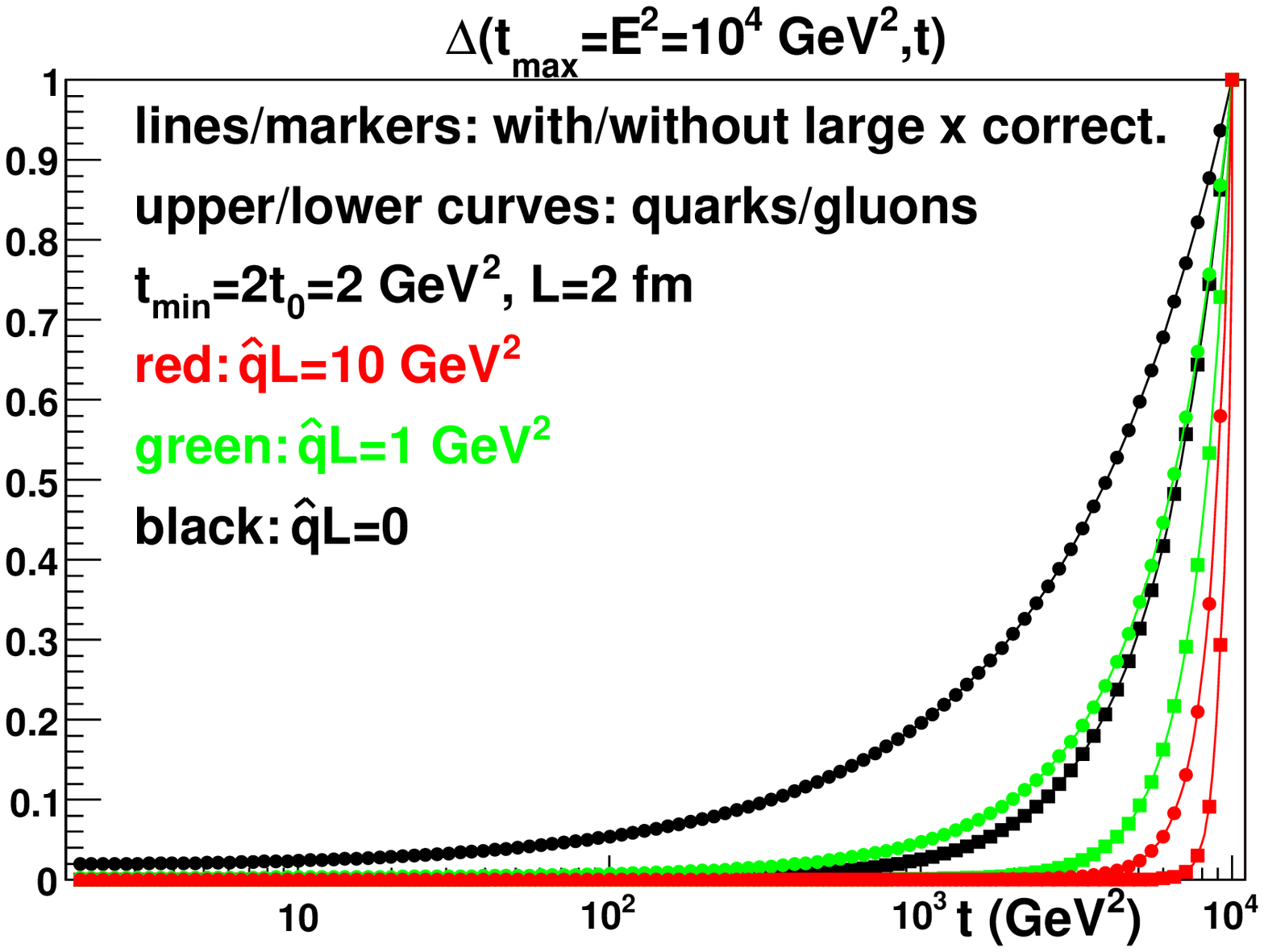}
\end{center}
\end{minipage}
\end{center}
\caption{Sudakov form factors for quarks (upper curves and symbols in each
plot) and
gluons (lower curves and symbols in each plot),
for parton energies $E=10$ GeV (upper plot) and $E=100$ GeV (lower plot), and for different medium parameters.}
\label{fig:sudakovs}
\end{figure}

The new splitting functions are also implemented in the calculation of the Sudakov form factors, the main component in the parton shower algorithms. The corresponding modifications are plotted in Fig. \ref{fig:sudakovs} where the vacuum and medium cases are compared for different values of the medium properties encoded in the jet quenching parameter $\hat q$. This comparison clearly shows the enhancement of the radiation expected by this new term in the splitting probability which will be translated into larger intrajet multiplicities and different jet structures.

In general we expect the jets produced by the modified parton shower to present: 
\begin{enumerate}

\item Larger multiplicities due to the enhancement in the radiation probability

\item  Softer spectra due to the energy loss of the partons by radiation, in particular that of the leading parton

\item Larger radiation angles (jet broadening) due to the absence of collinear divergencies in the medium radiation contribution. 

\end{enumerate}
These general expectations are independent of the particular implementation of the effect and are solely based on the properties of the medium-induced gluon radiation spectrum. A Monte Carlo implementation allows one to study the effect of important kinematic constrains which are otherwise difficult to address in analytical calculations.

In order to take into account the interplay between the medium length and the virtuality (or angular) ordering, a mixed approach is used in Ref. \cite{Armesto:2008qh,Armesto:2009fj}. For each splitting, the formation time of the gluon is computed as
\begin{equation}
t_{\rm form}\simeq\frac{2\omega}{k_\perp^2}
\label{eq:formtime}
\end{equation}
and the length in the splitting function (\ref{medsplit}) for the next branching corrected for this quantity. This effectively imposes a veto for radiation during the time in which a branching is being formed. This time is taken from the average value (\ref{eq:formtime}). So, although the ordering variable is the same as in the vacuum, some medium-radiation is vetoed by these formation time arguments.

The modified splitting probability and the corresponding corrections for length evolution are implemented as modifications of the shower routines in PYTHIA and HERWIG and available for public use \cite{Armesto:2008qh,Armesto:2009fj,prog,qatmc-site}. These are not official releases of the corresponding codes but just modifications of them. The nicknames Q-PYTHIA and Q-HERWIG are used for the corresponding implementations \cite{Armesto:2008qh,Armesto:2009fj,prog}.

\begin{figure}[h]
\begin{center}
\includegraphics[width=0.7\textwidth]{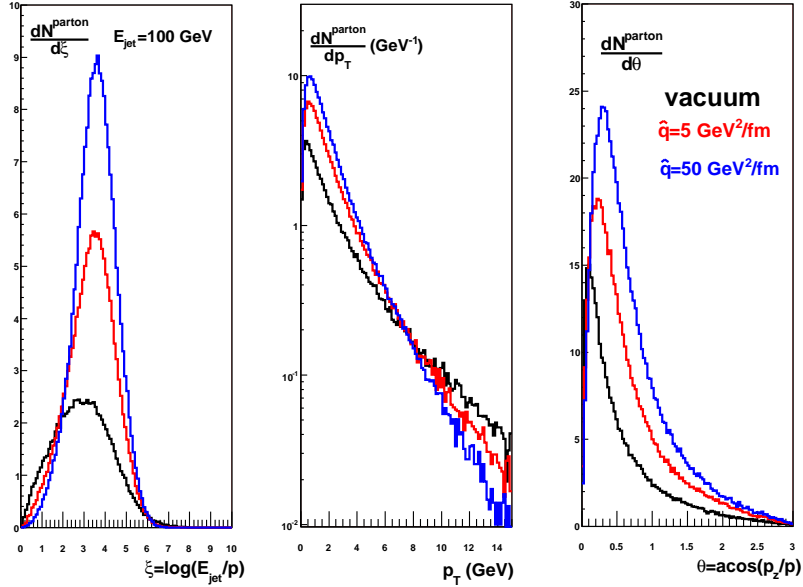}
\end{center}
\caption{Intrajet parton distributions in $\xi=\log(E_{\rm jet}/p)$ (left), $p_{T}$ (middle),
 and $\theta$ (right) for a gluon of initial energy $E_{jet}=100\, \mbox{GeV}$ in a medium of
 length  $L=2$ fm and for different transport coefficients $\hat{q}=0$ (black), 5 (red), and 50 (blue lines) GeV$^2$/fm.}
\label{fig:qpythia}       
\end{figure}

The intrajet parton distributions computed with Q-PYTHIA are plotted at the parton level (before hadronization) in Fig. \ref{fig:qpythia} for the fragmentation function (as a function of $\xi=-\log(z)$); the transverse momentum with respect to the jet axis; and the angle with respect to the jet axis. The main expectations listed above, the softening and enhancement of the multiplicity and the jet broadening, are clearly visible in this plot.

\section{The parton propagation in matter and the medium-induced gluon radiation}

In this section we provide general ideas on the calculation of the medium-induced gluon radiation and some of the properties of the spectrum. This spectrum has been computed by several groups \cite{Baier:1996sk,Zakharov:1998sv,Gyulassy:2000fs,Wiedemann:2000za,Arnold:2002ja} using different formalisms and approximations. For a more complete derivation, we refer the reader to the available lectures \cite{CasalderreySolana:2007zz} and reviews \cite{Baier:2000mf,Kovner:2003zj,Gyulassy:2003mc}, or the original papers.

\begin{figure}[h]
\begin{center}
\includegraphics[width=0.7\textwidth]{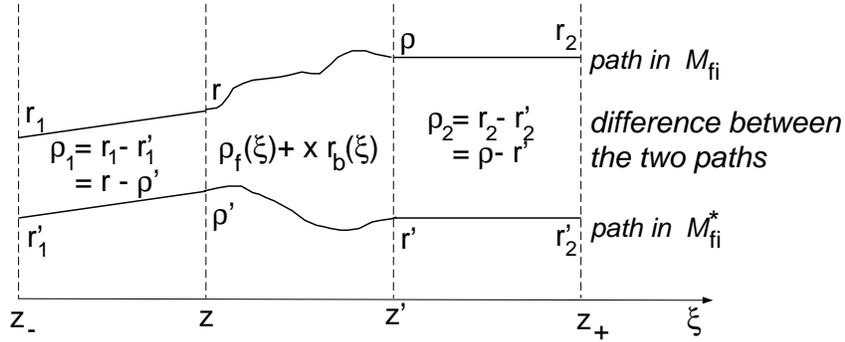}
\end{center}
\caption{Two paths as described by Eq. (\protect\ref{eq:pathint}). Figure from Ref. \cite{Wiedemann:1999fq}.}
\label{fig:urspath}       
\end{figure}

To describe the jet quenching phenomenon, we start from a high-$p_\perp$ quark or gluon produced in an elementary hard collision which subsequently interacts with the surrounding matter. So, the first question we need to address is how a highly energetic particle propagates through a dense medium. In the case of very energetic particles, a semiclassical approach is possible in which changes in the medium configuration due to the passage of the fast particle, recoil, are neglected. In these conditions, the propagation of a particle at transverse position ${\bf x}$ to transverse position ${\bf y}$ from time $t_1$ to $t_2$ can be described by the Feynman path integrals with a background field $A(x)$ (see Fig. \ref{fig:urspath}):
\begin{equation}
G({\bf x},t_1;{\bf y},t_2)=\int {\cal D}{\bf r}(t)\exp\left\{i\frac{E}{2}\int d\xi\left[\frac{d{\bf r}(\xi)}{d\xi}\right]^2+ig\int d\xi A({\bf r}(\xi))\right\}\, .
\label{eq:pathint}
\end{equation}
%

In the asymptotic case, $E\to \infty$, only the paths in which $d{\bf r}/d\xi=0$ survive, i.e., the particle does not change its transverse position, but travels in a straight line. In this case, the propagation is given by the Wilson line
\begin{equation}
W({\bf x})=\exp\left\{ig\int d\xi A({\bf x})\right\}.
\label{eq:wilson}
\end{equation}

Now, the propagation of a quark or a gluon through QCD matter (described by the fields $A(x)$) changes the phase of its wave function by rotation in colour space. 

To continue, we notice that any observable will involve colourless combinations of the Wilson lines, which implies that, at least, the colour trace of two Wilson lines should be preset for the colour to cancel. This colourless object will, then, {\it measure} the state of the medium in a given configuration. To compute a given observable, an average on the ensemble of possible medium configurations needs to be performed. So, the simplest object which will appear in any calculation is
 \begin{equation}
\frac{1}{N^2-1} {\rm Tr} \langle W({\bf x})W({\bf y}) \rangle\, ,
\label{eq:medav}
\end{equation}
where the prefactor $1/(N^2-1)$, with $N$ the number of colours, has been included to average over the initial colour configurations --- in this case for a gluon, for a quark the prefactor would be $1/N$. This object contains all the non-perturbative physics and, in the calculations presented here, only combinations such as ({\ref{eq:medav}) appear. 

One of the main issues in this type of analysis is to compute the medium averages (\ref{eq:medav}) for which several prescriptions exist. For a medium in which a large number of scattering centers interact with the propagating parton, an {\it opaque} medium, a widely used approximation is
 \begin{equation}
\frac{1}{N^2-1} {\rm Tr} \langle W({\bf x})W({\bf y}) \rangle\simeq\exp\left\{-\frac{1}{4}\int d\xi \hat q(\xi)({\bf x-y})^2\right\}\, .
\label{eq:qhatdef}
\end{equation}
This defines the transport coefficient, $\hat q$, as the prefactor of the typical small-distance ${\bf r}^2$-dependence of QCD dipole cross-sections. This prescription corresponds to propagating partons which describe Brownian motion in the transverse plane characterized by $\hat q$. The transport coefficient can, hence, be interpreted as the average transverse momentum squared per mean free path:
\begin{equation}
\hat q\simeq \frac{\langle k_T^2\rangle}{\lambda}\, .
\label{eq:qhatap}
\end{equation}

For the jet quenching calculations in the previous sections, we need to compute the medium-induced gluon radiation. This implies the propagation particle with large energy $E$, let us say a quark, which radiates a soft gluon with energy $\omega\ll E$ at a small angle $\theta\simeq k_\perp/\omega$. The typical diagram to be computed  can be seen in Fig. \ref{fig:medind} where the blobs represent scattering with the background field.
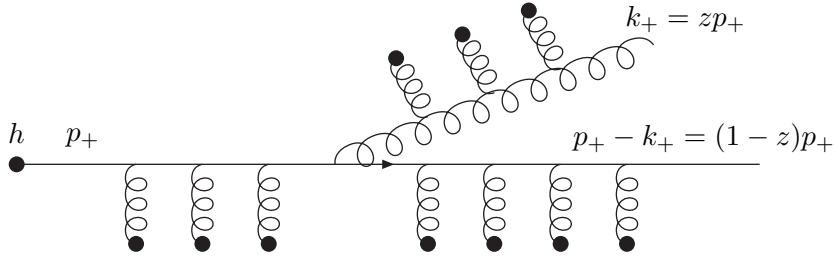
\begin{figure}
\begin{center}
\begin{picture}(300,90)(0,0)
\ArrowLine(10,30)(290,30)
\Gluon(55,0)(55,30){4}{3}
\Gluon(80,0)(80,30){4}{3}
\Gluon(105,0)(105,30){4}{3}
\Gluon(130,30)(250,75){5}{10}
\Gluon(165,0)(165,30){4}{3}
\Gluon(190,0)(190,30){4}{3}
\Gluon(215,0)(215,30){4}{3}
\Gluon(240,0)(240,30){4}{3}
\Gluon(165,48)(153,70){4}{3}
\Gluon(190,57)(178,79){4}{3}
\Gluon(215,66)(203,88){4}{3}
\Vertex(55,0){3}
\Vertex(80,0){3}
\Vertex(105,0){3}
\Vertex(165,0){3}
\Vertex(190,0){3}
\Vertex(215,0){3}
\Vertex(240,0){3}
\Vertex(153,70){3}
\Vertex(178,79){3}
\Vertex(203,88){3}
\Vertex(10,30){3}
\Text(10,42)[]{$h$}
\Text(35,40)[]{$p_+$}
\Text(270,40)[]{$p_+-k_+=(1-z)p_+$}
\Text(263,85)[]{$k_+=zp_+$}
\end{picture}
\caption{The medium-induced gluon radiation diagram}
\label{fig:medind}
\end{center}
\end{figure}
Figure \ref{fig:medind} contains the propagation of three particles in the medium, the initial quark and the produced gluon and quark. Each of these propagations are described by (\ref{eq:pathint}) --- in fact, the quark is considered completely eikonal, $E\to \infty$, so that the Wilson line (\ref{eq:wilson}) describes the propagation. In these conditions, the only non-perturbative object is the average of the Wilson line (\ref{eq:medav}). The final result takes a very compact form
\begin{eqnarray}
\omega\frac{dI}{d\omega d^2{\bf k_\perp}}=\frac{\alpha_S C_R}{(2\pi)^2\omega}2{\rm Re}\int_{x_{0}}^{L+x_0} \hspace{-0.35cm} dx\int d^2{\bf x}\ e^{-i{\bf k_\perp\cdot x}} \Bigg[\frac{1}{\omega}\int_{x}^{L+x_0}\hspace{-0.35cm} d\bar x\ e^{-\frac{1}{2}\int_{\bar x}^{L} d\xi n(\xi) \sigma({\bf x})}\times\nonumber\\ 
\times \frac{\partial}{\partial{\bf y}}\cdot\frac{\partial}{\partial{\bf x}}{\cal K}({\bf y}=0,x;{\bf x},\bar x)
-2\frac{\bf k_\perp}{{\bf k}_\perp^2}\cdot \frac{\partial}{\partial {\bf y}}{\cal K}({\bf y}=0,x;{\bf x},{L})\Bigg]+\frac{\alpha_S C_R}{\pi^2}\frac{1}{{\bf k}_\perp^2}\, ,
\label{eq:MIGRov}
\end{eqnarray}
where
\begin{equation}
{\cal K}\left({\bf r}(x),x;{\bf r}(\bar x),\bar x|\omega\right)=\int {\cal D}{\bf r}\exp\left[i\frac{\omega}{2} \int_{x}^{\bar x}d\xi\left(\left[\frac{d\bf{r}}{d\xi}\right]^2+i\frac{\hat q(\xi)}{2 \omega} {\bf r}^2\right)\right]
\label{eq:kpropap}
\end{equation}
corresponds to a 2-dimensional harmonic oscillator with time-dependent imaginary frequency. In Eq. ({\ref{eq:MIGRov}) the three terms correspond to (i) the gluon emitted inside the medium in both amplitude and conjugate amplitude; (ii) the emission inside the medium in amplitude and outside the medium in conjugate amplitude; (iii) and when the gluon is emitted outside the medium in both amplitude and conjugate amplitude (see Fig. \ref{fig:medind2}).
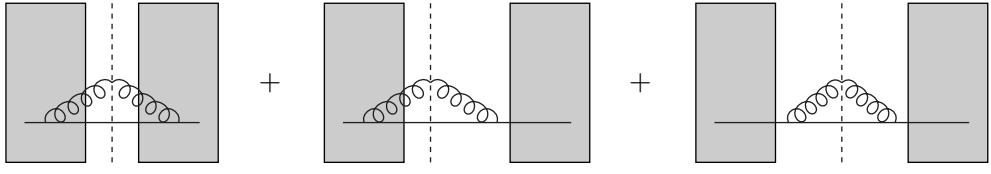
\begin{figure}
\begin{center}
\begin{picture}(350,90)(0,0)
\GBoxc(10,40)(30,60){0.8}
\GBoxc(60,40)(30,60){0.8}
\GBoxc(130,40)(30,60){0.8}
\GBoxc(200,40)(30,60){0.8}
\GBoxc(270,40)(30,60){0.8}
\GBoxc(350,40)(30,60){0.8}
\Line(2,25)(68,25)
\Gluon(10,25)(35,40){3}{4}
\Gluon(35,40)(60,25){3}{4}
\Line(122,25)(208,25)
\Gluon(130,25)(155,40){3}{4}
\Gluon(155,40)(180,25){3}{4}
\Line(262,25)(358,25)
\Gluon(290,25)(310,40){3}{4}
\Gluon(310,40)(330,25){3}{4}
\Text(95,40)[]{$+$}
\Text(235,40)[]{$+$}
\DashLine(35,10)(35,70){2}
\DashLine(155,10)(155,70){2}
\DashLine(310,10)(310,70){2}
\end{picture}
\caption{The three contributions to the squared amplitude of the medium-induced gluon radiation. The dashed line is the cut indicating the final outgoing particles.}
\label{fig:medind2}
\end{center}
\end{figure}

The last contribution corresponds to the vacuum radiation which is normally subtracted to define the medium-induced gluon radiation as
\begin{equation}
\omega\frac{dI}{d\omega d^2{\bf k_\perp}}=\omega\frac{dI^{\rm med}}{d\omega d^2{\bf k_\perp}}+\omega\frac{dI^{\rm vac}}{d\omega d^2{\bf k_\perp}}\, .
\label{eq:medraddef}
\end{equation}
The spectrum (\ref{eq:MIGRov}) with the vacuum subtraction (\ref{eq:medraddef}) is the building block of the calculations presented in Section \ref{sec:jetquenching}.

\subsection{Heuristic discussion}

In Fig. \ref{fig:static} we present the results for the double-differential medium-induced gluon radiation spectrum for a quark traversing a static medium of length $L$. The results are given as a function of the variables
\begin{equation}
\omega_c\equiv\frac{1}{2}\hat q\, L^2\, ,{\hspace{20pt}} \kappa^2\equiv\frac{k_\perp^2}{\hat qL}\, .
\label{eq:dimvar}
\end{equation}
%
\begin{figure}
\begin{minipage}{0.5\textwidth}
\begin{center}
\includegraphics[width=0.85\textwidth]{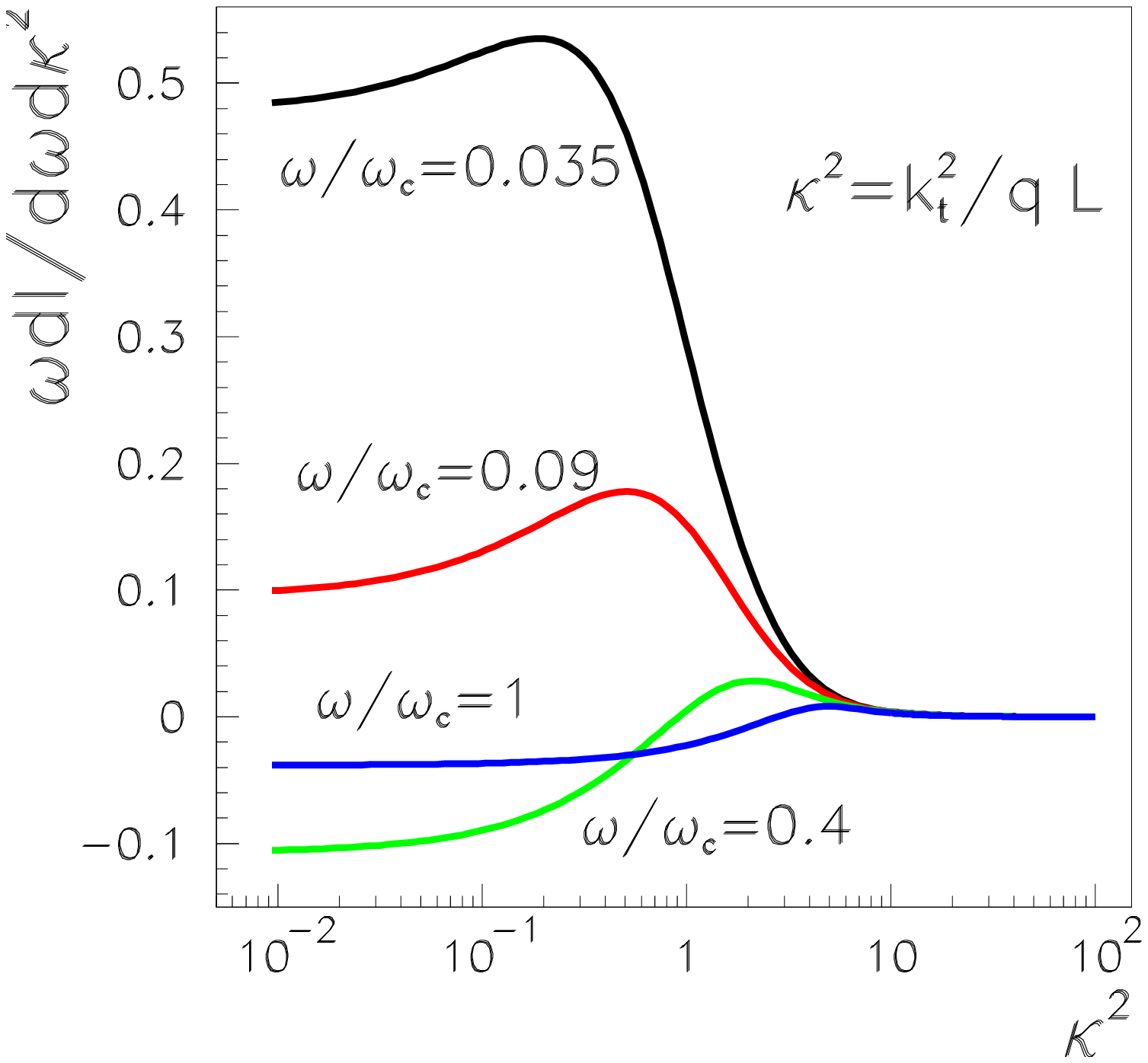}
\end{center}
\end{minipage}
\hfill
\begin{minipage}{0.5\textwidth}
\begin{center}
\includegraphics[width=0.85\textwidth]{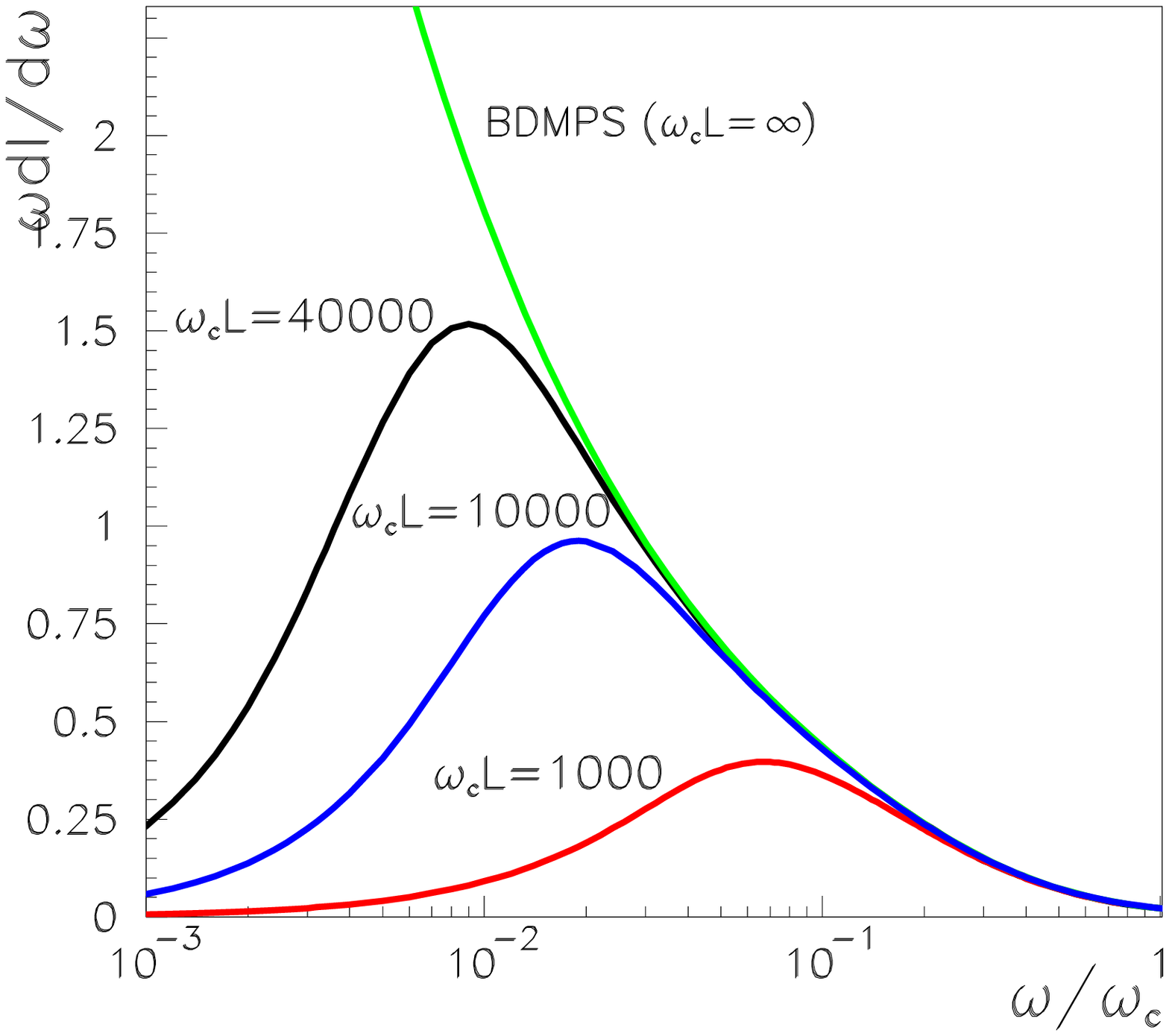}
\end{center}
\end{minipage}
\caption{Left: numerical results for the medium induced gluon radiation spectrum $\omega dI^{\rm med}/d\omega dk_\perp^2$ [Eqs. (\protect\ref{eq:MIGRov}) and (\protect\ref{eq:medraddef})] of a quark in a static medium as a function of the dimensionless variables (\ref{eq:dimvar}). Right: Same but integrated in $k_\perp<\omega$}
\label{fig:static}
\end{figure}
%
One important feature of the spectrum is the presence of small-$k_\perp$ and large-$\omega$ cuts which can be understood by the formation time of the gluon,
\begin{equation}
t_{\rm form}\simeq \frac{2\omega}{k_\perp^2}\, ,
\label{eq:form}
\end{equation}
which is controlled by the dynamical term in the path integral (\ref{eq:pathint}). To see this we can expand the path integral in the number of scatterings
\begin{eqnarray}
G({\bf x},t_0;{\bf y},t_f)=G_0(t_0\to t_f)+G_0(t_0\to t_1)igA({\bf x_1})G_0(t_1\to t_f)+\nonumber\\ 
+\frac{1}{2}G_0(t_0\to t_1)igA({\bf x_1})G_0(t_1\to t_2)igA({\bf x_2})G_0(t_2\to t_f)\dots
\label{eq:opexp}
\end{eqnarray}
where $G_0(t_i\to t_j)$ is a short-cut notation for the vacuum propagator from time $t_i$ to $t_j$ with transverse positions defined by ${\bf x_i}$ and ${\bf x_j}$, respectively. The vacuum propagator can be written as the Fourier transform of a plane wave
\begin{equation}
G_0({\bf x_{i\perp}},t_i;{\bf x_{(i+1)\perp}},t_{(i+1)})=
\int \frac{d^2p_{i\perp}}{(2\pi)^2}\ {\rm e}^{i\frac{p_{i\perp}^2}{2\omega}((t_{i}-t_{(i+1)})}
{\rm e}^{-i{\bf p_{i\perp}}({\bf x_{i\perp}-x_{(i+1)\perp}})}\, .
\label{eq:vacprop}
\end{equation}
In (\ref{eq:vacprop}) the exponential factors define the coherence length (\ref{eq:form}): when $t_{\rm form} \gg t_i-t_{(i+1)}$ the phase is small and we can write
\begin{equation}
\prod_i\int \frac{d^2p_{i\perp}}{(2\pi)^2}\ 
{\rm e}^{-i{\bf p_{i\perp}}({\bf x_{i\perp}-x_{(i+1)\perp}})}\to\int d{\bf x_\perp}{\rm e}^{-i{\bf p_{i\perp}}({\bf x}-{\bf y})}\, .
\end{equation}
So, all the scatterings (\ref{eq:opexp}) take part at the same time --- coherently. On the opposite side, when the formation time is much smaller than typical lengths in the medium, the exponentials oscillate very fast and only the first term survives. In other words, in the incoherent case, the cross-section {\it counts} the number of scattering centers, while in the coherent case the whole medium acts as a single scattering center. As a result, a reduction of the gluon radiation is produced in the latter case. This is the generalization to QCD of the Landau--Pomeranchuk--Migdal effect. The numerical effect appears clearly in Fig. \ref{fig:static} as a suppression of the spectrum for small values of $\kappa^2$. An important consequence is that the spectrum is neither collinearly divergent (i.e., it can be safely integrated to $k_\perp=0$) nor infrared divergent (i.e., it can be integrated to $\omega=0$) as can be seen in Fig. \ref{fig:static}. In contrast the vacuum part of the spectrum (\ref{eq:MIGRov}) presents both collinear and soft divergences,
\begin{equation}
\frac{dI^{\rm vac}}{d\omega d^2k_\perp}=\frac{\alpha_s C_R}{\pi^2}\frac{1}{k_\perp^2}\frac{1}{\omega}\, .
\end{equation}
The position of the infrared cut in the $k_\perp$-integrated spectrum (Fig. \ref{fig:static} Right) can also be understood in terms of the formation time: integrating the spectrum in the kinematically allowed region $0<k_t<\omega$, and noticing\footnote{This can be estimated by taking $\langle k^2_t\rangle\sim \hat q t_{\rm form}$, using Eq. (\protect\ref{eq:form}) $\langle k^2_t\rangle\sim\sqrt{\hat q\omega}$.} that $\langle k^2_t\rangle\sim\sqrt{\hat q \omega}$ then a suppression of the spectrum for $\omega\lesssim \hat q^{1/3}$ should appear.

\subsection*{Acknowledgements}

I would like to thank the organizers of {\it The 2008 European School of High-Energy Physics} for the invitation to this nice meeting. Special thanks to N\'estor Armesto for a critical reading of the manuscript. 
This work is supported by CICYT of Spain under the projects FPA2005-01963 and FPA2008-01177;
by the Spanish Consolider--Ingenio 2010 Programme CPAN (CSD2007-00042);
by Xunta de Galicia grant INCITE08PXIB296116PR, and
by the European Commission grant PERG02-GA-2007-224770.
C.A. Salgado is a Ram\'on y Cajal researcher.

\end{document}